\documentclass[aps,pra,reprint,twocolumn,nofootinbib,superscriptaddress]{revtex4-2}
\usepackage{epsfig,amssymb,amsmath,physics,enumitem,qcircuit,bm}
\usepackage{color,subfig}
\graphicspath{{./fig/}}

\usepackage{graphicx,ragged2e}
\usepackage{etoolbox}
\makeatletter
\renewcommand{\@makecaption}[2]{%
  \vskip\abovecaptionskip
  \sbox\@tempboxa{#1: #2}%
  \ifdim \wd\@tempboxa > \hsize
    \begin{center} 
      \small\justifying #1: #2
    \end{center}%
  \else
    \begin{center} 
      \small #1: #2
    \end{center}%
  \fi
  \vskip\belowcaptionskip}
\makeatother

\definecolor{myurlcolor}{rgb}{0,0,0.7}
\definecolor{myrefcolor}{rgb}{0.8,0,0}
\usepackage{hyperref}
\hypersetup{colorlinks,linkcolor=blue,anchorcolor=blue,citecolor=blue,urlcolor=blue}

\newtheorem{theorem}{Theorem}

\newtheorem{lemma}{Lemma}

\newtheorem{corollary}{Corollary}

\def\equationautorefname~#1\null{%
  Eq.~(#1)\null
}
\def\figureautorefname~#1\null{%
  FIG.~#1\null
}
\def\subfigureautorefname~#1\null{%
  FIG.~#1\null
}
\def\sectionautorefname~#1\null{%
  Sec.~#1\null
}

\renewcommand{\a}{\hat{a}}
\renewcommand{\b}{\hat{b}}
\newcommand{\ad}{\hat{a}^\dag}
\newcommand{\bd}{\hat{b}^\dag}
\newcommand{\n}{\hat{n}}
\newcommand{\q}{\hat{q}}
\newcommand{\p}{\hat{p}}
\newcommand{\Q}{\hat{Q}}
\renewcommand{\P}{\hat{P}}

\newcommand{\C}{\mathbb{C}}
\newcommand{\R}{\mathbb{R}}
\newcommand{\Z}{\mathbb{Z}}

\newcommand{\D}[1]{\mathcal{D}\left(#1\right)}
\renewcommand{\SS}[1]{\mathcal{S}\left(#1\right)}
\newcommand{\RR}[1]{\mathcal{R}\left(#1\right)}
\newcommand{\BS}[1]{\mathcal{BS}\left(#1\right)}
\newcommand{\CD}[1]{\mathcal{CD}\left(#1\right)}
\newcommand{\CR}[1]{\mathcal{CR}\left(#1\right)}
\newcommand{\CS}[1]{\mathcal{CS}\left(#1\right)}
\newcommand{\CBS}[1]{\mathcal{CBS}\left(#1\right)}

\newcommand{\F}{\mathcal{F}}
\newcommand{\M}{\mathcal{M}}
\newcommand{\Par}{\mathcal{P}}
\newcommand{\SWAP}{\mathcal{SWAP}}
\newcommand{\CP}{\mathcal{CP}}

\newcommand{\tF}{\widetilde{\mathcal{F}}}
\newcommand{\tPar}{\widetilde{\mathcal{P}}}
\newcommand{\tSWAP}{\widetilde{\mathcal{SWAP}}}
\newcommand{\tCP}{\widetilde{\mathcal{CP}}}
\newcommand{\CNOT}{\mathrm{CNOT}}

\newcommand{\encq}{\mathrm{Enc}_{Q}}
\newcommand{\encp}{\mathrm{Enc}_{P}}

\newcommand{\ketf}[1]{\ket{#1}_{F}}

\newcommand{\hf}[1]{\frac{#1}{2}}
\newcommand{\hfth}{\frac{\theta}{2}}
\newcommand{\qtth}{\frac{\theta}{4}}
\newcommand{\sqhf}[1]{\frac{#1}{\sqrt{2}}}
\newcommand{\inv}{^{-1}}
\newcommand{\eps}{\epsilon}

\newcommand{\tpsi}{\tilde{\psi}}
\newcommand{\tj}{\tilde{j}}
\newcommand{\tk}{\tilde{k}}
\newcommand{\ts}{\tilde{s}}

\newcommand{\bmq}{\bm{q}}
\newcommand{\bmp}{\bm{p}}
\newcommand{\bmj}{\bm{j}}

\newcommand{\pp}[1]{\left(#1\right)}
\newcommand{\bb}[1]{\left[#1\right]}
\newcommand{\kk}[1]{\left\{#1\right\}}

\renewcommand{\pmat}[1]{\begin{pmatrix} #1 \end{pmatrix}}

\newcommand{\cip}[2]{{}_C\!\ip{#1}{#2}}

\newcommand{\sgn}{\mathrm{sgn}}

\newcommand{\pf}[1]{\textit{Proof of \autoref{#1}.}}
\newcommand{\qed}{\hfill$\square$}
\newcommand{\appref}[1]{Appx.~\ref{#1}}
\newcommand{\NCSUECE}{Department of Electrical and Computer Engineering,\\ North Carolina State University, Raleigh, NC 27606, USA}
\newcommand{\NCSUCS}{Department of Computer Science, North Carolina State University, Raleigh, NC 27606, USA}
\newcommand{\NCSUPh}{Department of Physics and Astronomy, North Carolina State University, Raleigh, NC 27606, USA}
\newcommand{\NCSUMATH}{Department of Mathematics, North Carolina State University, Raleigh, NC 27606, USA}

\usepackage{orcidlink}

\begin{document}

\title{Efficient Qubit Simulation of Hybrid Oscillator-Qubit Quantum Computation}

\author{Xi Lu}
\affiliation{\NCSUECE}


\author{Bojko N. Bakalov}
\affiliation{\NCSUMATH}

\author{Yuan Liu\orcidlink{0000-0003-1468-942X}}
\thanks{q\_yuanliu@ncsu.edu}
\affiliation{\NCSUECE}
\affiliation{\NCSUCS}
\affiliation{\NCSUPh}

\begin{abstract}
    We introduce a framework for simulating hybrid oscillator-qubit quantum processors on qubit-only systems through position encoding.
    By encoding continuous-variable position and momentum wave functions into qubit amplitudes, our method efficiently simulates all Gaussian and conditional Gaussian operations -- encompassing the phase-space instruction set (beam splitter, single-qubit rotation, conditional displacement) and extending to squeezing, conditional squeezing, conditional rotation, and conditional beam splitter -- using $O(\log^2(\Gamma + \log(\epsilon\inv)))$ qubit gates per hybrid gate, where $\Gamma$ is the Fock-level bound and $\epsilon$ is the target precision.
    This polylogarithmic per-gate complexity represents an exponential improvement over Fock basis encoding approaches, which require exponential quantum or classical resources in the number of qubits per mode.
    We provide rigorous numerical characterization of quantum Fourier transform errors for Fock-bounded states, enabling precise resource estimation for practical implementations.
    This work establishes that hybrid oscillator-qubit algorithms can be implemented on qubit processors with polynomial overhead, providing new insights into the computational power trade-offs between discrete-variable and hybrid continuous-discrete-variable quantum computing.
\end{abstract}

\maketitle

\section{Introduction}\label{sec:intro}

Quantum computation has largely focused on discrete-variable (DV, qubit) systems.
In parallel, continuous-variable (CV, bosonic oscillator) systems have been studied for decades~\cite{glauber1951some,glauber1963coherent,marian1993squeezed,hamilton2017gaussian,zhong2020quantum,madsen2022quantum,dutta2025simulating}, offering unique advantages due to their infinite-dimensional Hilbert spaces.
CV systems provide robust encodings including GKP codes~\cite{gottesman2001encoding,terhal2020towards}, cat codes~\cite{mirrahimi2014dynamically,ofek2016extending,lee2024fault,hillmann2023quantum}, and binomial codes~\cite{michael2016new,albert2018performance}.

Hybrid CV-DV systems have attracted significant attention for balancing the easy calibration of DV systems with the resource efficiency of CV systems~\cite{liu2024hybrid,crane2024hybrid,eickbusch2022fast}.
These architectures show promise for quantum simulation~\cite{schweizer2019floquet}, quantum link models~\cite{chandrasekharan1997quantum}, Schwinger models~\cite{shaw2020quantum}, error correction~\cite{grimsmo2021quantum,gottesman2001encoding}, quantum signal processing~\cite{sinanan2024single,low2017optimal,martyn2021grand}, and simulation of bosonic and fermionic systems~\cite{crane2024hybrid,kang2023leveraging}.

Understanding the computational power of these hybrid systems requires characterizing the role of non-Gaussian resources.
It is well established that Clifford-only computation can be efficiently simulated classically; achieving universal quantum computation requires non-Clifford resources, termed magic~\cite{howard2014contextuality,pashayan2015estimating}.
Analogously, non-Gaussian operations are essential for universal CV quantum computation~\cite{walschaers2021non,jacobs2007engineering,teoh2023dual} and constitute the primary challenge for classical simulation.
The conditional displacement gate forms the \emph{phase-space instruction set}~\cite{eickbusch2022fast} with single-qubit rotations and beam splitters, while the Jaynes-Cummings gate forms the \emph{sideband instruction set}~\cite{liu2021constructing,mischuck2013qudit} with single-qubit rotations and displacements.
Bosonic circuits without non-Gaussian resources can be efficiently simulated classically~\cite{barthe2025gatebased}, and computational advantages have been characterized through non-Gaussian stellar ranks~\cite{chabaud2022holomorphic,chabaud2020stellar,chabaud2023resources}.

This invites a key question: How much computational power do hybrid CV-DV systems possess in comparison to pure DV systems?
This parallels investigations comparing qubit and qudit systems~\cite{jankovic2024noisy,keppens2025qudit}.
Since CV states exist in infinite-dimensional Hilbert spaces, one can truncate to finite dimensions and simulate on qubits up to a specified error tolerance~\cite{tong2022provably,arzani2025can}.
However, finite-dimensional truncations of infinite-dimensional quantum systems require careful treatment, as truncated Hamiltonians may converge to incorrect dynamics depending on the choice of basis~\cite{fischer2026quantum}.
Moreover, recent work has explored the trade-offs between system size and energy requirements in hybrid oscillator-qubit models for quantum computation~\cite{brenner2025trading}.

Most existing approaches rely on Fock basis encoding~\cite{crane2024hybrid,tong2022provably,arzani2025can}, which imposes a Fock-level cutoff and maps CV modes to DV states by encoding Fock states into qubit computational basis states.
Under this scheme, ladder operators $\hat{a}, \hat{a}^\dagger$ can be simulated using linear combinations of unitaries~\cite{simon2025ladder} or quantum signal processing~\cite{liu2025efficient}.
However, simulating elementary hybrid CV-DV gates---including displacement, beam splitter, and conditional beam splitter~\cite{liu2024hybrid,eickbusch2022fast}---via Fock basis encoding requires quantum or classical resources exponential in $n$, the number of qubits per mode.

\begin{table*}
    \centering
    \begin{tabular}{|c|c|c|c|}
    \hline
    \textbf{Hybrid CV-DV Gate}
    & \textbf{DV Circuit}
    & \textbf{Gate Count}
    & \textbf{Error Bound}
    \\ \hline
    Displacement $\D{\alpha}=e^{\alpha\ad-\alpha^*\a}$
    & \autoref{eq:dv-disp}
    & $2$ QFT, $2n$ $R_z$
    & $2\eps_F$
    \\ \hline
    Rotation $\RR{\theta}=e^{-i\theta(\n+\hf{1})}$
    & \autoref{eq:dv-rotation}
    & $2$ QFT, $\frac{3n(n-1)}{2}$ $C_z$
    & $2\eps_F$
    \\ \hline
    Squeezing $\SS{r}=e^{\hf{r}(\a^2-\a^{\dag 2})}$
    & \autoref{eq:dv-squeezing}
    & $4$ QFT, $2n(n-1)$ $C_z$
    & $4\eps_F$
    \\ \hline
    Beam Splitter $\BS{\theta}=e^{-i\hf{\theta} (\ad\b+\a\bd)}$
    & \autoref{eq:dv-bs}
    & $4$ QFT, $3n^2$ $C_z$
    & $4\eps_F$
    \\ \hline
    Conditional Displacement $\CD{\alpha}=e^{Z(\alpha\ad-\alpha^*\a)}$
    & \autoref{eq:dv-cd}
    & $2$ QFT, $2n$ $R_z$, $2n$ CNOT
    & $2\eps_F$
    \\ \hline
    Conditional Rotation $\CR{\theta}=e^{-i\theta Z(\n+\hf{1})}$
    & \autoref{eq:dv-cr}
    & $2$ QFT, $\frac{3n(n-1)}{2}$ $C_z$, $6n$ CNOT
    & $2\eps_F$
    \\ \hline
    Conditional Squeezing $\CS{r}=e^{\hf{r}Z(\a^2-\a^{\dag 2})}$
    & \autoref{eq:dv-cs}
    & $4$ QFT, $4n(n-1)$ $C_z$, $8n$ CNOT
    & $4\eps_F$
    \\ \hline
    Conditional Beam Splitter $\CBS{\theta}=e^{-i\hf{\theta} Z(\ad\b+\a\bd)}$
    & \autoref{eq:dv-cbs}
    & $4$ QFT, $3n^2$ $C_z$, $6n$ CNOT
    & $4\eps_F$
    \\ \hline
    Heterodyne Detection
    & \autoref{fig:dv-heterodyne}
    & $4$ QFT, $3n^2$ $C_z$
    & $4\eps_F$
    \\ \hline
    Photon Number Counting
    & \autoref{fig:dv-photon-counting}
    & $2\gamma$ QFT, $\frac{3\gamma n(n-1)}{2}$ $C_z$, $6\gamma n$ CNOT
    & $2\gamma\eps_F$
    \\ \hline
    \end{tabular}
    \caption{
        Simulation of hybrid CV-DV gates on DV systems using position encoding with $n$ qubits per mode ($N=2^n$ grid points, spacing $\lambda=\sqrt{2\pi/N}$).
        Each QFT costs $n$ H, $\frac{n(n-1)}{2}$ $R_z$, and $n(n-1)$ CNOT gates (\autoref{tab:basic-gates}).
        Error bounds are expressed as $k\eps_F$ where $k$ is the number of QFTs and $\eps_F$ is the QFT error of the intermediate states (\autoref{cor:qft-error-general}).
        For photon number counting, $\gamma=\lceil\log_2(\Gamma+1)\rceil$.
    }
    \label{tab:gate-decomp}
\end{table*}

\begin{table*}
    \centering
    \begin{tabular}{|c|c|c|}
        \hline
        \textbf{Elementary Hybrid Gate}
        & \textbf{Qubit Simulation Circuit}
        & \textbf{Simulation Cost}
        \\ \hline
        $\F$ and $\F\inv$
        & QFT and inverse
        & $n$ H, $\hf{n(n-1)}$ $R_z$, $n(n-1)$ CNOT
        \\ \hline
        $\Par$
        & $\sigma_x$ on each qubit
        & $n$ X
        \\ \hline
        $\SWAP$
        & swap each qubit pair
        & $n$ SWAP
        \\ \hline
        $e^{it\q}$
        & \autoref{eq:enc-disp-q}
        & $n$ $R_z$
        \\ \hline
        $e^{it\q_1\q_2}$
        & \autoref{eq:e-jk}
        & $n^2$ $R_z$, $2n^2$ CNOT
        \\ \hline
        $e^{it\q^2}$
        & \autoref{eq:e-jj}
        & $\hf{n(n-1)}$ $R_z$, $n(n-1)$ CNOT
        \\ \hline
        $e^{itZ\q_1\q_2}$
        & \autoref{eq:enc-cejk}
        & $n^2$ $R_z$, $2n^2+2n$ CNOT
        \\ \hline
        $\CP$
        & \autoref{eq:dv-cp}
        & $n$ CNOT
        \\ \hline
        $e^{itZ\q}$
        & \autoref{eq:dv-cd}
        & $n$ $R_z$, $2n$ CNOT
        \\ \hline
        $e^{itZ\q^2}$
        & \autoref{eq:enc-cejj}
        & $\hf{n(n-1)}$ $R_z$, $n(n+1)$ CNOT
        \\ \hline
        Homodyne Detection
        & $M_z$
        & $n$ $M_z$
        \\ \hline
    \end{tabular}
    \caption{
        Simulation cost of elementary gates using position encoding with $n$ qubits per mode.
        The left column lists CV gates; the right columns give the corresponding qubit simulation circuits and costs.
        Gate notation: H: Hadamard; $R_z$: Z-rotation; CNOT: controlled-NOT; $M_z$: Pauli-Z measurement.
    }
    \label{tab:basic-gates}
\end{table*}

Alternatively, CV systems can be discretized using wave function (quadrature, position or momentum) basis representations~\cite{jordan2012quantum,hastrup2022universal,macridin2022bosonic,macridin2024qumode}.
With this encoding, displacement gates can be simulated efficiently, though rotations and beam splitters lack straightforward simulation methods, as they do not have simple unitary forms in discretized position or momentum bases.
Field amplitude discretization was first proposed in~\cite{jordan2012quantum} for simulating $\phi^4$ quantum field theory, with gate complexity scaling as $(1/\epsilon)^{O(1)}$.
Recent work~\cite{macridin2022bosonic,macridin2024qumode} developed a more precise discretization from the perspective of Nyquist--Shannon sampling, establishing a framework for discretizing momentum operators using the quantum Fourier transform (QFT) and implementing squeezing with polynomial resources~\cite{macridin2024qumode}. The computational complexity of preparing Gaussian states on qubit registers has been well studied~\cite{kitaev2008wavefunction,Rattew2021efficient,bauer2021practical,kuklinski2025simpler}, as this is essentially equivalent to preparing a CV Gaussian state under position encoding. However, extensions to operator simulation and hybrid CV-DV systems have not been established.
An alternative Fock-state-based approach uses the quantum Hermite transform to map computational basis states to harmonic oscillator eigenstates, enabling efficient quantum circuits for high-dimensional $SU(n)$ representations~\cite{iyer2026efficient}.

In this paper, we study the simulation of hybrid CV-DV systems on pure DV systems using position encoding.
We demonstrate that hybrid CV-DV operations with at most quadratic Hamiltonians can be simulated using $O(\log^2(\Gamma + \log(\epsilon\inv)))$ elementary gates per hybrid gate, where $\Gamma$ is the Fock-level bound and $\epsilon$ is the target precision.
This encompasses the phase-space instruction set~\cite{eickbusch2022fast} (beam splitter, single-qubit rotation, and conditional displacement), and extends beyond it to include squeezing, conditional squeezing, conditional rotation, and conditional beam splitter---covering all Gaussian and conditional Gaussian operations.
Our key contributions include:
(1) efficient DV simulation circuits for all Gaussian and conditional Gaussian operations with rigorous error bounds;
(2) systematic numerical characterization of quantum Fourier transform errors enabling precise resource estimation; and
(3) explicit demonstration that operations with at most quadratic Hamiltonians require only polylogarithmic overhead, representing an exponential improvement over Fock basis approaches.
\autoref{tab:gate-decomp} and \autoref{tab:basic-gates} summarize our main technical results.

\section{State Encoding}\label{sec:enc}

In this section, we establish the mathematical framework for encoding CV states into DV states, applicable to both pure CV and hybrid CV-DV systems.

We use $\ketf{\cdot}$ to denote the Fock states. We use the convention $\a=\sqhf{\q+i\p}$, where $\a,\q,\p$ are the annihilation operator, position operator, and momentum operator, respectively. The canonical operators $\{\q_1,\ldots,\q_m;\p_1,\ldots,\p_m\}$ satisfy the canonical commutation relations $[\q_j,\q_k] = [\p_j,\p_k] = 0$ and $[\q_j,\p_k] = i\delta_{jk}$.

We use wave function (quadrature) encoding, which discretizes the position or momentum representation of CV states~\cite{jordan2012quantum,macridin2022bosonic,macridin2024qumode}. Given an $m$-mode state $\ket{\psi}$ with position wave function $\psi(\bmq) = \psi(q_1,\cdots,q_m)$ and momentum wave function $\tpsi(\bmp)=\tpsi(p_1,\cdots,p_m)$, we choose $n$ qubits per mode, set $N=2^n$, and define the spacing parameter $\lambda=\sqrt{2\pi/N}$, which we will see is the natural choice to balance position and momentum discretization.

The \emph{position encoding} samples the position wave function on a grid:
\begin{equation}\label{eq:enc-q}
\begin{aligned}
    \encq\pp{\ket{\psi}} := \lambda^{m/2}
    \sum_{\bmj\in[N]^{m}}
    \psi(\lambda\tilde{\bmj})
    \ket{\bmj},
\end{aligned}
\end{equation}
where $\bmj=(j_1,\cdots,j_m)$ is a multi-index, $[N]:=\kk{0,\cdots,N-1}$, $\tj := j - (N-1)/2$ is the shifted index, and $\ket{\bmj}=\ket{j_1}\cdots\ket{j_m}$ is a computational basis state on $mn$ qubits (organized as $m$ registers of $n$ qubits each).

The \emph{momentum encoding} samples the momentum wave function analogously:
\begin{equation}\label{eq:enc-p}
\begin{aligned}
    \encp\pp{\ket{\psi}} := \lambda^{m/2}
    \sum_{\bmj\in[N]^{m}}
    \tpsi(\lambda\tilde{\bmj})
    \ket{\bmj}.
\end{aligned}
\end{equation}
In the continuous setting, the position and momentum wave functions are related by the Fourier transform:
\begin{equation}\label{eq:cft}
\begin{aligned}
    \tpsi(\bmp) = (2\pi)^{-m/2} \int_{\R^m} \psi(\bmq) e^{-i\bmp\cdot\bmq} \dd\bmq.
\end{aligned}
\end{equation}
On DV systems, we use the shifted quantum Fourier transform (QFT) to switch between position and momentum encodings:
\begin{equation}\label{eq:def-qft}
\begin{aligned}
    \tF \ket{j} := \frac{1}{\sqrt{N}} \sum_{k=0}^{N-1} e^{-i\lambda^2\tj\tk} \ket{k}, \quad (j=0,\cdots,N-1).
\end{aligned}
\end{equation}
The choice of spacing $\lambda=\sqrt{2\pi/N}$ ensures that the QFT phase factor $\lambda^2 = 2\pi/N$ matches the standard QFT, and that position and momentum grids have equal spacing---a natural choice that balances discretization in both bases.
The key property is that $\tF \encq(\ket{\psi}) \approx \encp(\ket{\psi})$ with error determined by aliasing of wave function tails outside the computational window.

The fundamental principle underlying our simulation framework is that any unitary containing only position operators $\{\q_j\}$ is exact on position encoding, and any unitary containing only momentum operators $\{\p_j\}$ is exact on momentum encoding.
The only source of error in simulating hybrid CV-DV gates arises from quantum Fourier transforms (QFT) that switch between position and momentum representations.
More precisely, for a unitary of the form $U = e^{iH(\q_1,\ldots,\q_m)}$ depending only on position operators, the DV simulation satisfies $\encq(U\ket{\psi}) = \encq(U) \encq(\ket{\psi})$, meaning the simulation is exact up to initial encoding error.
Similarly, $e^{iH(\p_1,\ldots,\p_m)}$ is exact on momentum encoding.

As a concrete example, consider the rotation gate $\RR{\theta}=e^{-i\theta(\n+\hf{1})}$.
From \appref{app:decomp-rot}, the rotation can be decomposed as
\begin{equation}\label{eq:rotation-preview}
    \RR{\theta} = e^{-\hf{i}\tan\hfth\q^2} \cdot e^{-\hf{i}\sin\theta\p^2} \cdot e^{-\hf{i}\tan\hfth\q^2},
\end{equation}
which consists of position-only operations (exact on position encoding), momentum-only operations (exact on momentum encoding), and QFTs to switch between bases.
Similarly, squeezing $\SS{r}$ decomposes into 4 elementary operations with 4 QFTs, and beam splitter $\BS{\theta}$ uses two-mode versions of these operations.
Full decompositions are given in \autoref{sec:gaussian}.

Our error analysis centers on the \emph{QFT error} $\eps_F$ arising from basis switching via quantum Fourier transform.
Since QFT is a discrete approximation of the continuous Fourier transform between position and momentum wave functions, it introduces truncation error.
The QFT can be implemented with $O(n^2)$ gates: $n$ Hadamard gates, $n(n-1)/2$ controlled-$R_z$ gates, and $n(n-1)$ CNOT gates (see \autoref{tab:basic-gates}).
We define the \emph{QFT error} of a CV state $\ket{\psi}$ as,
\begin{equation}\label{eq:def_qft_err}
\begin{aligned}
    &
    \eps_F(\ket{\psi}) := \norm{
        \encp{\ket{\psi}} - \tF \encq{\ket{\psi}}
    }
\end{aligned}
\end{equation}

For a general gate $U$, we define the \emph{gate discretization error} as the norm of the difference between the discretization of the continuous unitary applied to the original state and the discretized unitary applied to the discretized state:
\begin{equation}\label{eq:def_gate_err}
    \epsilon_U(\ket{\psi}) := \norm{
        \encq(U\ket{\psi}) - \encq(U)\encq(\ket{\psi})
    }.
\end{equation}
For gates containing only position operators, $\epsilon_U = 0$ on position encoding; similarly, momentum-only gates have zero error on momentum encoding. The only error source is QFT basis switching.

\begin{figure*}[t]
    \centering
    \includegraphics[width=.7\textwidth]{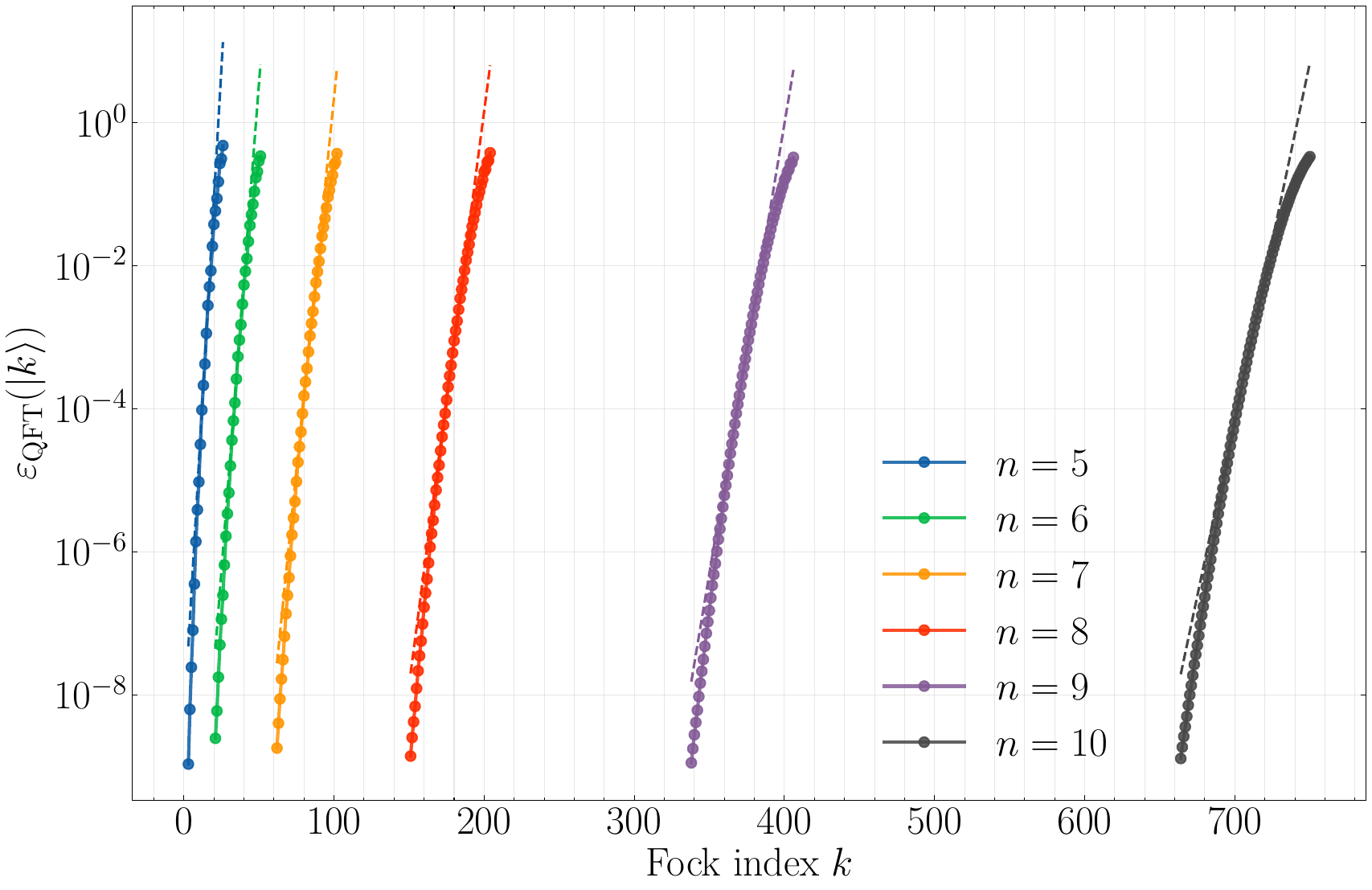}
    \caption{
        Per-Fock QFT error $\eps_F(\ketf{k})$ as a function of Fock level $k$ for different grid sizes $N = 2^n$.
        The linear relationship on the log scale confirms the per-Fock scaling $\log\eps_F(\ketf{k}) \lesssim ak + b$.
        The dashed lines show the fitted formula in \autoref{eq:qft-err-numerical} with coefficients in \autoref{eq:qft-coeff-scaling}.
    }
    \label{fig:qft-err-scaling}
\end{figure*}

We characterize the QFT error for individual Fock states through systematic numerical experiments.
We compute the error when applying the shifted QFT to each Fock state $\ketf{k}$ independently, comparing against the analytically known Fourier-transformed Fock state $(-i)^k\ketf{k}$.
The numerical results are shown in \autoref{fig:qft-err-scaling}.

\begin{lemma}[Per-Fock QFT Error]\label{lem:qft_err_per_fock}
For Fock states $\ketf{k}$ simulated with $N = 2^n$ grid points, the QFT error satisfies
\begin{equation}\label{eq:qft-err-numerical}
\begin{aligned}
    \eps_F(\ketf{k}) \lesssim e^{a_n k + b_n},
\end{aligned}
\end{equation}
where
\begin{equation}\label{eq:qft-coeff-scaling}
\begin{aligned}
    a_n &\approx 4.2125 \times 2^{-n/2} + 0.1027, \\
    b_n &\approx -5.7903 \times 2^{n/2} + 16.2724,
\end{aligned}
\end{equation}
for register size $5\le n\le 10$ and error range $10^{-9} \le \eps_F \le 10^{-1}$.
\end{lemma}

The exponential growth $\eps_F(\ketf{k}) \sim e^{a_n k}$ with Fock level $k$ is confirmed in \autoref{fig:qft-err-scaling}.

\begin{corollary}[General-State QFT Error]\label{cor:qft-error-general}
For any state $\ket{\psi} = \sum_{k} c_k \ketf{k}$ simulated with $N = 2^n$ grid points,
\begin{equation}\label{eq:qft-err-general}
    \eps_F(\ket{\psi}) \le \sqrt{\ev{e^{2a_n\n+2b_n}}},
\end{equation}
where $a_n,b_n$ are as in \autoref{lem:qft_err_per_fock} and $\ev{e^{2a_n\n+2b_n}} = \sum_k |c_k|^2 e^{2a_n k+2b_n}$.
The bound may exceed 1 at high Fock levels; however, for physically relevant states the amplitudes $|c_k|^2$ decay rapidly, so high-Fock contributions are negligible and the bound remains informative.
\end{corollary}

\pf{cor:qft-error-general} By the triangle inequality and \autoref{lem:qft_err_per_fock},
\begin{equation}
\begin{aligned}
    &
    \eps_F(\ket{\psi})
    \le 
    \sum_k |c_k|\,e^{a_n k+b_n}
    \\ \le &
    \sqrt{\sum_k |c_k|^2 e^{2a_n k+2b_n}}
    =
    \sqrt{\ev{e^{2a_n\n+2b_n}}}.
\end{aligned}
\end{equation}
\qed

The $O(n^2)$ gate cost per QFT combined with the error scaling enables complete resource budgeting.
For example, to achieve $\eps_F(\ketf{k}) \lesssim 10^{-6}$ for $k \le 10$, solving \autoref{eq:qft-err-numerical} yields $n = 6$ qubits per mode ($N = 64$ grid points).

\emph{Initial state encoding.}
In many applications the initial CV state is Gaussian---for instance, the vacuum or a coherent/thermal state.
Kitaev and Webb~\cite{kitaev2008wavefunction} give an algorithm that prepares an $m$-mode Gaussian with covariance matrix $\Sigma$ on $mn$ qubits in two stages: (i) prepare $m$ independent 1D Gaussians, each using a circuit whose size is polynomial in $n$ and $\log(1/\eps)$; (ii) apply $O(m^2)$ shearing operations---one per off-diagonal entry of the $LDL^T$ factor of $\Sigma^{-1}$---each costing $O(n^2)$ two-qubit gates, yielding $O(m^2 n^2)$ total gates for the correlation step~\cite{bauer2021practical}.
Thus the one-time encoding cost scales as $O(m^2 n^2)$, the same order as simulating $O(m^2)$ hybrid gates.
For a circuit with $K$ hybrid gates (typically $K \gg m^2$), the encoding overhead is negligible.
Simpler near-linear-in-$n$ single-mode preparation circuits (replacing step (i)) have also been proposed~\cite{kuklinski2025simpler}.

\begin{figure*}[t]
    \centering
    \subfloat[Displacement $\D{2}$ on $\ketf{k}$]{\includegraphics[width=0.48\textwidth]{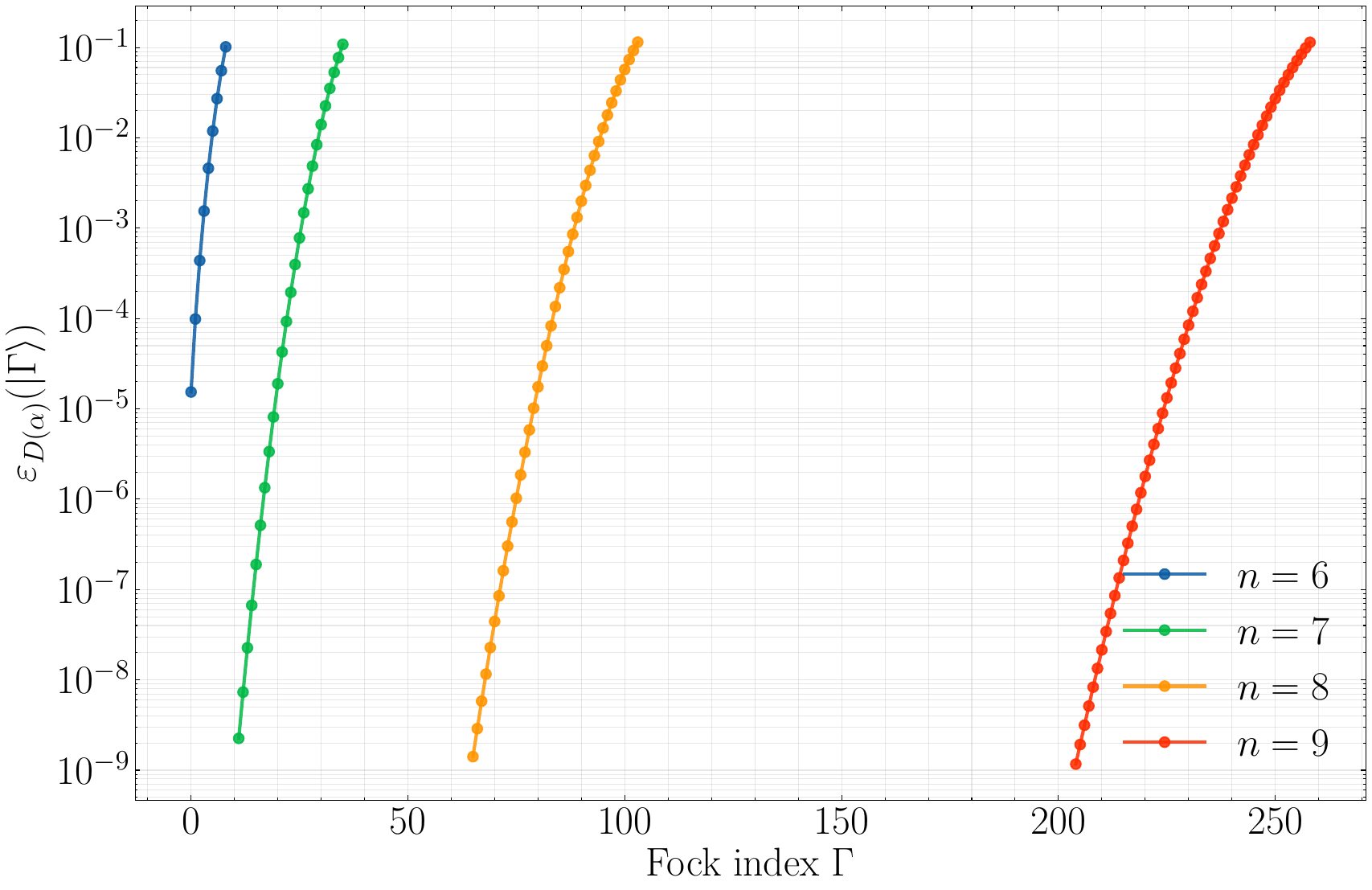}}
    \hfill
    \subfloat[Rotation $\RR{\pi/4}$ on $\ketf{k}$]{\includegraphics[width=0.48\textwidth]{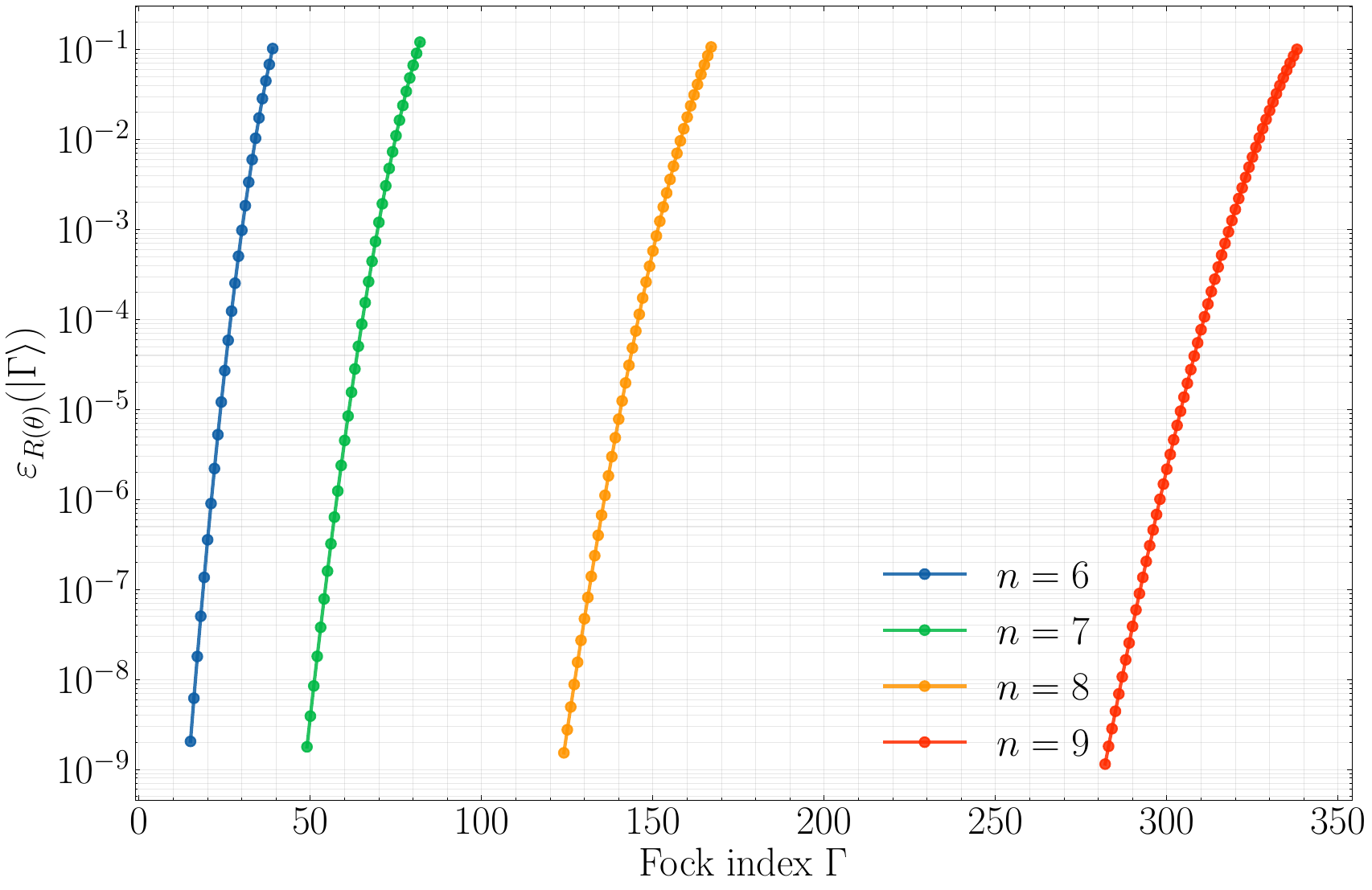}}
    \\
    \subfloat[Squeezing $\SS{1}$ on $\ketf{k}$]{\includegraphics[width=0.48\textwidth]{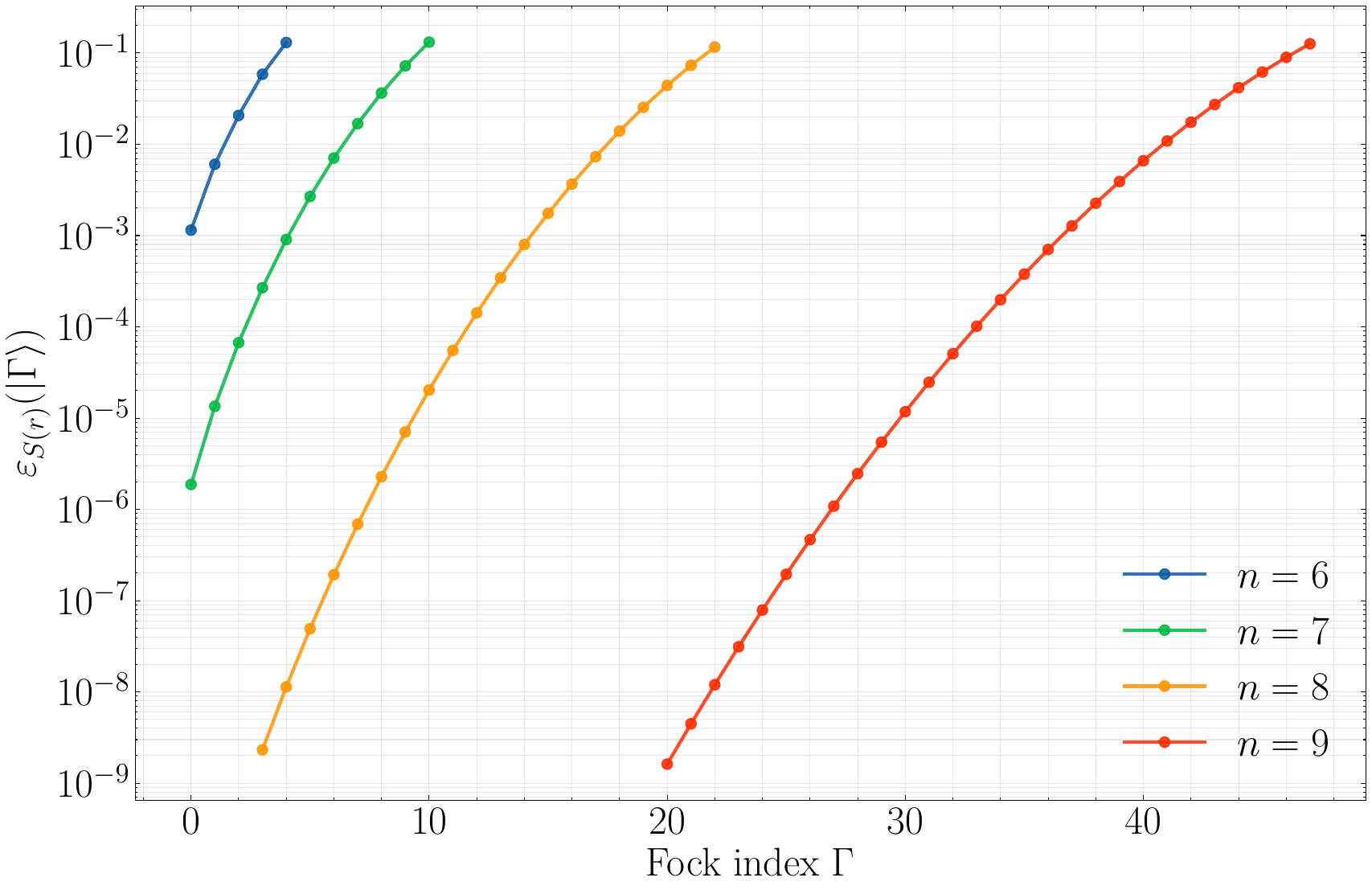}}
    \hfill
    \subfloat[Beam Splitter $\BS{\pi/2}$ on $\ketf{k}\ketf{0}$]{\includegraphics[width=0.48\textwidth]{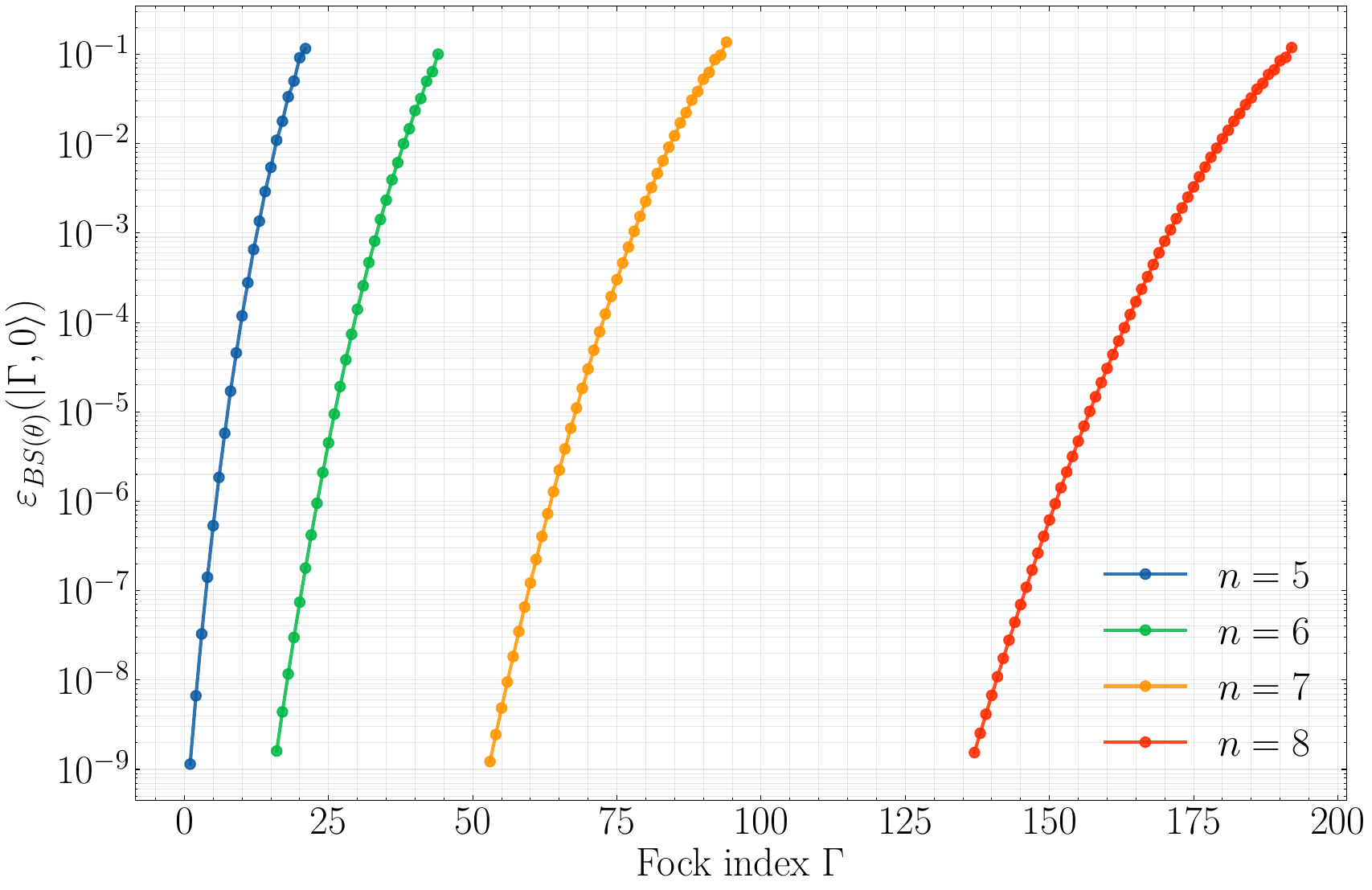}}
    \caption{
        Per-Fock gate discretization error $\epsilon_U(\ketf{k})$ as a function of Fock level $k$ for different grid sizes $N = 2^n$.
        Each point gives the error when the gate is applied to the individual Fock state $\ketf{k}$.
        (a) Displacement with $\alpha=2$.
        (b) Rotation with $\theta=\pi/4$.
        (c) Squeezing with $r=1$.
        (d) Beam splitter with $\theta=\pi/2$ applied to $\ketf{k}\ketf{0}$.
    }
    \label{fig:gate-errors}
\end{figure*}

\section{Oscillator Gates}\label{sec:gaussian}

In this section, we show how to simulate Gaussian operations on DV systems using our position encoding framework. The key result is that any Gaussian operation with a quadratic Hamiltonian can be decomposed into elementary operations of the form $e^{it\q_j\q_k}$, $e^{it\p_j\p_k}$, $e^{it\q_j\p_k}$ (\autoref{thm:basis-cv}), each of which is either exact on position/momentum encoding or requires QFT for basis switching.

\subsection{Displacement}

For notational convenience, we introduce shorthand notation for the discrete position and momentum operators:
\begin{equation}\label{eq:def-QP}
    \Q := \encq\pp{\q}, \quad \P := \tF^\dag \Q \tF.
\end{equation}

The displacement operator on a single mode is defined as,
\begin{equation}\label{eq:def-disp}
    \D{\alpha} = e^{\alpha\ad-\alpha^*\a}, \quad(\alpha\in\C),
\end{equation}
and can be decomposed as,
\begin{equation}\label{eq:disp-decomp}
\begin{aligned}
    \D{\alpha}
    = &
    e^{-i\Im\alpha\Re\alpha} \D{i\Im\alpha} \D{\Re\alpha}
    \\ = &
    e^{-i\Im\alpha\Re\alpha} e^{i\sqrt{2}(\Im\alpha)\q} e^{-i\sqrt{2}(\Re\alpha)\p}.
\end{aligned}
\end{equation}


The encoding of $\q$ and $e^{it\q}$ is defined as,
\begin{equation}
    \Q =
    \sum_{j=0}^{N-1} \lambda\tj \dyad{j}{j},
\end{equation}
where $\tj\equiv j-\hf{N-1}$, and
\begin{equation}\label{eq:enc-disp-q}
    \encq\pp{
        e^{it\q}
    } :=
    e^{it\Q}
    =
    \sum_{j=0}^{N-1} e^{it\lambda\tj} \ket{j}\bra{j}.
\end{equation}

\begin{figure}[t]
\centering
\includegraphics[width=.5\textwidth]{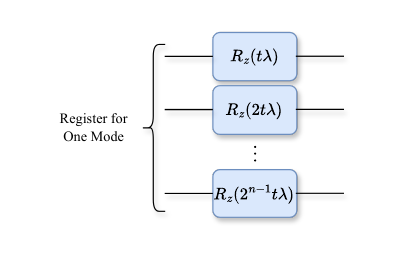}
\caption{
    Qubit circuit to simulate the elementary displacement operator $e^{it\q}$ on position-encoding qubits.
}
\label{fig:dv-disp-elem}
\end{figure}

One can implement $\encq\pp{e^{it\q}}$ by,
\begin{equation}
\begin{aligned}
    &
    \encq\pp{
        e^{it\q}
    }
    \\ =&
    \sum_{j_0,j_1,\cdots,j_{n-1}=0}^1
    e^{it\lambda\pp{
        j_0 + 2 j_1 + \cdots + 2^{n-1} j_{n-1} - \hf{N-1}
    }} \dyad{j}
    \\ =&
    \prod_{k=0}^{n-1}
    e^{-it 2^{k-1}\lambda Z_k},
\end{aligned}
\end{equation}
which consists of $n$ rotation-Z gates $R_z(\theta)=e^{i\hf{\theta} Z}$, where $Z_k$ is the Pauli-Z operator on the $k$-th qubit.
The circuit is shown in \autoref{fig:dv-disp-elem}.

The unitary error of $e^{it\q}$ on position-encoding qubits is always 0, since
\begin{equation}\label{eq:zero-err}
\begin{aligned}
    &
    \encq\pp{e^{it\q}\ket{\psi}}
    =
    \sqrt{\lambda} \sum_{j=0}^{N-1}
    e^{it\lambda\tj} \psi(\lambda\tj) \ket{j}
    \\=&
    \pp{
        \sum_{j=0}^{N-1}
        e^{it\lambda\tj} \dyad{j}{j}
    }
    \pp{
        \sqrt{\lambda} \sum_{k=0}^{N-1}
        \psi(\lambda\tk) \ket{k}
    }
    \\=&
    \encq\pp{e^{it\q}} \encq\pp{\ket{\psi}}.
\end{aligned}
\end{equation}

The exponential of the discrete momentum operator $\P$ (defined in \autoref{eq:def-QP}) is
\begin{equation}\label{eq:enc-disp-p}
    \encq\pp{
        e^{it\p}
    } :=
    e^{it\P}
    =
    \tF^\dag e^{it\Q} \tF.
\end{equation}

For the most general case $\alpha\in\C$, one first decomposes it with \autoref{eq:disp-decomp} and then simulates the two axis-aligned displacements separately.
If one wants to start and end in the position basis, then one QFT and one inverse QFT are needed to switch the basis.
The total cost is the sum of the costs of the two axis-aligned displacements, i.e., 2 QFTs plus $2n$ $R_z$ gates.

In summary, the simulation of a displacement operator $\D{\alpha}$ on position-encoding qubits can be done by the following circuit,
\begin{equation}\label{eq:dv-disp}
\begin{aligned}
    &
    \encq\pp{
        \D{\alpha}
    }
    \\ = &
    \encq\pp{
        e^{i\sqrt{2}(\Re\alpha) \q}
    }
    \tF^\dag \encp\pp{
        e^{-i\sqrt{2}(\Im\alpha) \p}
    } \tF
    \\ = &
    \pp{
        \prod_{j=0}^{n-1} e^{-i\sqrt{2}(\Re\alpha) (2^j\lambda)Z_j}
    }
    \tF^\dag \pp{
        \prod_{j=0}^{n-1} e^{i\sqrt{2}(\Im\alpha) (2^j\lambda)Z_j}
    } \tF .
\end{aligned}
\end{equation}

Since position-only and momentum-only operations are exact on their respective encodings (as shown in \autoref{eq:zero-err}), the unitary error comes solely from the 2 QFTs required for basis switching.
\begin{equation}\label{eq:disp-err-bound}
    \epsilon_{\D{\alpha}}(\ket{\psi}) \lesssim 2\eps_F.
\end{equation}

The gate cost of $\D{\alpha}$ is 2 QFTs plus $2n$ $R_z$ gates, totaling $O(n^2)$ elementary qubit gates.

\subsection{Gaussian Operators with Quadratic Hamiltonians}

In this section, we show how to simulate arbitrary quadratic Gaussian operations on DV systems.
We focus here on quadratic Gaussian operations other than displacement. Such operations can be expressed as a symplectic matrix $F$ acting on the quadrature basis in the Heisenberg picture; displacements---which introduce additional constant shifts---are already covered in the previous section.

In the Heisenberg picture, quadratic Gaussian operations are transformations on the quadrature basis $\hat{\bm{r}}=\pp{\q_1,\cdots,\q_m;\p_1,\cdots,\p_m}^\top$,
\begin{equation}\label{eq:gauss-sp}
    \hat{\bm{r}}'
    =
    F \hat{\bm{r}},
\end{equation}
for some symplectic matrix $F$.


\begin{theorem}[Elementary Decomposition]\label{thm:basis-cv}
    Any Gaussian operation with a quadratic Hamiltonian on a CV system can be expressed as a product of the following operators:
    \begin{equation}\label{eq:elem-quad}
    \begin{gathered}
        E_{j,k}^{(1)}(t) = e^{it\q_j\q_k}, \quad
        E_{j,k}^{(2)}(t) = e^{it\p_j\p_k}, \quad
        E_{j,k}^{(3)}(t) = e^{it\q_j\p_k}, \\
        (1\le j \ne k\le m);
        \\
        E_{j,j}^{(1)}(t) = e^{it(\q_j)^2}, \quad
        E_{j,j}^{(2)}(t) = e^{it(\p_j)^2}, \quad
        (1\le j \le m).
    \end{gathered}
    \end{equation}
\end{theorem}

\pf{thm:basis-cv}
Since,
\begin{align}
    E_{j,k}^{(1)\dag}(t) \q_s E_{j,k}^{(1)}(t) &= \q_s, \\
    E_{j,k}^{(1)\dag}(t) \p_s E_{j,k}^{(1)}(t) &= \p_s + t (\delta_{j,s}\q_k + \delta_{k,s}\q_j),
\end{align}
the symplectic matrix corresponding to $E_{j,k}^{(1)}$ is,
\begin{equation}
    E_{j,k}^{(1)}(t) \Leftrightarrow \begin{pmatrix}
        I & 0 \\
        t (e_{jk}+e_{kj}) & I
    \end{pmatrix},
\end{equation}
where $e_{jk}$ is the $m\times m$ matrix with $1$ at $(j,k)$ and $0$ elsewhere.
Similarly, the symplectic matrices corresponding to $E_{j,k}^{(2)}$ and $E_{j,k}^{(3)}$ are,
\begin{align}
    E_{j,k}^{(2)}(t) &\Leftrightarrow \begin{pmatrix}
        I & -t (e_{jk}+e_{kj}) \\
        0 & I
    \end{pmatrix},
    \\
    E_{j,k}^{(3)}(t) &\Leftrightarrow \begin{pmatrix}
        I - t e_{jk} & 0 \\
        0 & I + t e_{kj}
    \end{pmatrix},
\end{align}
respectively.
All the matrices above are called \emph{elementary symplectic matrices}~\cite{o1978symplectic}, and by Sec.~2.2 of~\cite{o1978symplectic}, they generate the symplectic group $Sp(2m,\R)$, i.e., any symplectic matrix can be expressed as a product of elementary symplectic matrices.

\qed


We note that this is not the usual way to implement a Gaussian operation on analog hardware, but it provides a systematic decomposition suited for qubit simulation.

The elementary Gaussian gate $E_{j,k}^{(1)}(t)=e^{it\q_j\q_k}$ can be discretized on position encoding as,
\begin{equation}\label{eq:e-jk}
\begin{aligned}
    &
    \encq\pp{e^{it\q_j\q_k}}
    \\=& \sum_{r,s=0}^{N-1} e^{it(\lambda\tilde{r})(\lambda\ts)} \dyad{r}{r}_{j} \otimes \dyad{s}{s}_{k}
    \\=&
    \prod_{r,s=0}^{n-1} e^{i2^{r+s-1}\lambda^2 t Z_{j,r} Z_{k,s}},
\end{aligned}
\end{equation}
where $Z_{j,r}$ is the Pauli-Z operator on the $r$-th qubit of the qubit register that simulates the $j$-th mode, and $\tF_{j}$ (or $\tF_{j,k}$) is the quantum Fourier transform on the qubit register that simulates the $j$-th mode (or the $j$-th and $k$-th modes).
Other elementary Gaussian gates can be encoded similarly on corresponding position/momentum encoding qubits.

\begin{figure*}[t]
\centering
\subfloat[
    Qubit circuit to simulate the elementary Gaussian operator $e^{it\q_j\q_k}$ on position-encoding qubits.
]{
    \includegraphics[width=.9\linewidth]{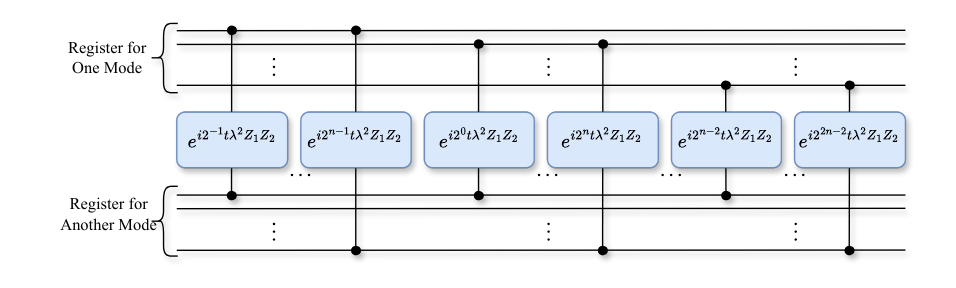}
    \label{fig:dv-ejk-elem}
}
\\
\subfloat[
    Qubit circuit to simulate the elementary Gaussian operator $e^{it\q_j^2}$ on position-encoding qubits.
]{
    \includegraphics[width=.9\linewidth]{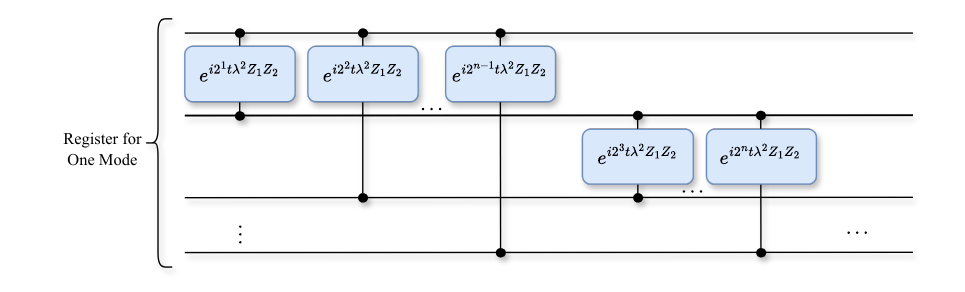}
    \label{fig:dv-ejj-elem}   
}
\caption{
    Qubit circuits to simulate elementary quadratic Gaussian operators.
}
\end{figure*}

The encoding of $e^{it\q_j\q_k}(j\ne k)$ consists of $n^2$ conditional Z-rotation gates $C_z(\theta):=e^{-i\hfth Z_1Z_2}$.
When $j=k$, \autoref{eq:e-jk} can be simplified to,
\begin{equation}\label{eq:e-jj}
    \encq\pp{e^{it\q_j^2}} = \prod_{0\le r < s \le n-1} e^{i2^{r+s}\lambda^2 t Z_{j,r} Z_{j,s}},
\end{equation}
which costs $\hf{n(n-1)}$ $C_z$ gates.
The circuits are shown in \autoref{fig:dv-ejk-elem} and \autoref{fig:dv-ejj-elem}.

\begin{figure*}[t]
    \centering
    \subfloat[$\tSWAP$]{\includegraphics[width=0.28\textwidth]{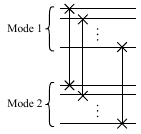}\label{fig:dv-swap}}
    \hfill
    \subfloat[$\tPar$]{\includegraphics[width=0.36\textwidth]{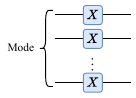}\label{fig:dv-parity}}
    \hfill
    \subfloat[$\tCP$]{\includegraphics[width=0.28\textwidth]{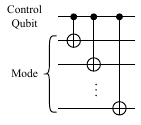}\label{fig:dv-cp}}
    \caption{The (a) $\tSWAP$, (b) $\tPar$, and (c) $\tCP$ gates, each corresponding to the discretization of their continuous counterparts, $\SWAP$, $\Par$, and $\CP$, respectively.}
\end{figure*}

In \appref{app:decomp}, we show the decomposition of rotation $\RR{\theta}=e^{-i\theta(\n+\hf{1})}$, squeezing $\SS{r}=e^{\hf{r}(\a^2-\a^{\dag 2})}$, and beam splitter $\BS{\theta}=e^{-i\hf{\theta}(\ad\b+\a\bd)}$.
The results are as follows.
The encoding of the rotation $\RR{\theta}$ on position-encoding qubits is,
\begin{equation}\label{eq:dv-rotation}
\begin{aligned}
    &
    \encq\pp{\RR{\theta}}
    \\ =&
    e^{-\hf{i}\tan\hfth\Q^2} e^{-\hf{i}\sin\theta\P^2} e^{-\hf{i}\tan\hfth\Q^2}
    \\ =&
    \pp{
        \prod_{0\le s < t < n} e^{-i2^{s+t-1}\lambda^2 \tan\hfth Z_s Z_t}
    }
    \\ &
    \tF^\dag
    \pp{
        \prod_{0\le s < t < n} e^{i2^{s+t}\lambda^2 \sin\theta Z_s Z_t}
    }
    \tF
    \\ &
    \pp{
        \prod_{0\le s < t < n} e^{-i2^{s+t-1}\lambda^2 \tan\hfth Z_s Z_t}
    }
\end{aligned}
\end{equation}
when $\abs{\theta}\le\frac{\pi}{4}$, and
\begin{equation}\label{eq:dv-rotation-general}
\begin{aligned}
    &
    \encq\pp{\RR{\theta}} = \encq\pp{\RR{\theta_0}} \encq\pp{\RR{\theta-\theta_0}},
\end{aligned}
\end{equation}
where $\theta_0\in\hf{\pi}\Z$ is chosen such that $\abs{\theta-\theta_0}\le\frac{\pi}{4}$, and
\begin{equation}\label{eq:dv-large-rot}
\begin{aligned}
    \encq\pp{\RR{\hf{\pi}}} =& \tF,
    \\
    \encq\pp{\RR{\pi}} =& \tPar,
    \\
    \encq\pp{\RR{-\hf{\pi}}} =& \tF^\dag,
\end{aligned}
\end{equation}
where $\tF$ is the discretized Fourier transform (see \autoref{eq:def-qft}), and $\tPar = \prod_{j=0}^{n-1} X_{j}$ is the discretized parity gate (\autoref{fig:dv-parity}).

The simulation error for rotation is determined by the QFT errors at each basis-switching step.
The decomposition uses 2 QFTs, giving
\begin{equation}\label{eq:rot-err-bound}
    \eps_{\RR{\theta}}(\ket{\psi}) \lesssim 2\eps_F.
\end{equation}
The per-Fock error $\eps_{\RR{\pi/4}}(\ketf{k})$ is shown in \autoref{fig:gate-errors}(b).

The gate cost of $\RR{\theta}$ is 2 QFTs plus 3 shearing layers $e^{it\q^2}$, each costing $\frac{n(n-1)}{2}$ $C_z$ gates, totaling $O(n^2)$ elementary qubit gates.

The encoding of the squeezing $\SS{r}$ on position-encoding qubits is,
\begin{equation}\label{eq:dv-squeezing}
\begin{aligned}
    &
    \encq\pp{\SS{r}}
    \\ =&
    e^{i\hf{\mu_4(r)}\Q^2}
    e^{i\hf{\mu_3(r)}\P^2}
    e^{i\hf{\mu_2(r)}\Q^2}
    e^{i\hf{\mu_1(r)}\P^2},
    \\ =&
    \pp{
        \prod_{0\le s < t < n} e^{i2^{s+t-1}\lambda^2 \mu_4(r) Z_s Z_t}
    }
    \\ &
    \tF^\dag
    \pp{
        \prod_{0\le s < t < n} e^{i2^{s+t-1}\lambda^2 \mu_3(r) Z_s Z_t}
    }
    \tF
    \\ &
    \pp{
        \prod_{0\le s < t < n} e^{i2^{s+t-1}\lambda^2 \mu_2(r) Z_s Z_t}
    }
    \\ &
    \tF^\dag
    \pp{
        \prod_{0\le s < t < n} e^{i2^{s+t-1}\lambda^2 \mu_1(r) Z_s Z_t}
    }
    \tF,
\end{aligned}
\end{equation}
where
\begin{equation}\label{eq:mu-coeffs}
\begin{aligned}
    \mu_1(r) &= -e^{r/2}\sqrt{\abs{e^{r}-1}}, \\
    \mu_2(r) &= \sgn(r)e^{-r/2}\sqrt{\abs{e^{r}-1}}, \\
    \mu_3(r) &= e^{-r/2}\sqrt{\abs{e^{r}-1}}, \\
    \mu_4(r) &= -\sgn(r)e^{r/2}\sqrt{\abs{e^{r}-1}},
\end{aligned}
\end{equation}
where $\sgn(r)=1$ if $r\ge 0$ and $\sgn(r)=-1$ if $r<0$.

The squeezing operation $\SS{r}$ requires 4 QFTs in its decomposition, giving
\begin{equation}\label{eq:sq-err-bound}
    \eps_{\SS{r}}(\ket{\psi}) \lesssim 4\eps_F.
\end{equation}
The per-Fock error $\eps_{\SS{1}}(\ketf{k})$ is shown in \autoref{fig:gate-errors}(c).

The gate cost of $\SS{r}$ is 4 QFTs plus 4 shearing layers $e^{it\q^2}$, each costing $\frac{n(n-1)}{2}$ $C_z$ gates, totaling $O(n^2)$ elementary qubit gates.
Compared with \cite{macridin2024qumode}, which decomposes the squeezing gate into 6 elementary gates using a Trotter-based approach, our symplectic elementary decomposition reduces this to 4 elementary gates.

The encoding of the beam splitter $\BS{\theta}$ on position-encoding qubits is,
\begin{equation}\label{eq:dv-bs}
\begin{aligned}
    &
    \encq\pp{\BS{\theta}}
    \\ =&
    e^{-\hf{i}\tan\qtth\Q_1\Q_2}
    e^{-\hf{i}\sin\hfth\P_1\P_2}
    e^{-\hf{i}\tan\qtth\Q_1\Q_2}
    \\ =&
    \pp{
        \prod_{r,s=0}^{n-1} e^{-i2^{r+s-1}\lambda^2 \tan\qtth Z_{1,r} Z_{2,s}}
    }
    \\ &
    \tF_1^\dag \tF_2^\dag
    \pp{
        \prod_{r,s=0}^{n-1} e^{-i2^{r+s-1}\lambda^2 \sin\hfth Z_{1,r} Z_{2,s}}
    }
    \\ &
    \tF_1 \tF_2
    \pp{
        \prod_{r,s=0}^{n-1} e^{-i2^{r+s-1}\lambda^2 \tan\qtth Z_{1,r} Z_{2,s}}
    },
\end{aligned}
\end{equation}
when $\abs{\theta}\le\frac{\pi}{2}$.
For other angles, one can decompose it as $\BS{\theta}=\BS{\theta_0}\BS{\theta-\theta_0}$, where $\theta_0\in\pi\Z$ and $\abs{\theta-\theta_0}\le\frac{\pi}{2}$.
The encoding of the beam splitter $\BS{\theta_0}$ is given by,
\begin{equation}\label{eq:dv-bs-theta0}
\begin{aligned}
    &\encq\pp{\BS{\pi}} = \tF_1 \tF_2 \tSWAP,
    \\
    &\encq\pp{\BS{2\pi}} = \tPar_1 \tPar_2,
    \\
    &\encq\pp{\BS{-\pi}} = \tF_1^\dag \tF_2^\dag \tSWAP,
\end{aligned}
\end{equation}
where $\tSWAP$ consists of one swap gate between the $j$-th qubits of the two mode registers for each $j$, as in \autoref{fig:dv-swap}.

The beam splitter operation $\BS{\theta}$ requires 4 QFTs (2 per mode) in its decomposition, giving
\begin{equation}\label{eq:bs-err-bound}
    \eps_{\BS{\theta}}(\ket{\psi}) \lesssim 4\eps_F.
\end{equation}
The per-Fock error $\eps_{\BS{\pi/2}}(\ketf{k}\ketf{0})$ (total photon number $k$) is shown in \autoref{fig:gate-errors}(d).

The gate cost of $\BS{\theta}$ is 4 QFTs plus 3 shearing layers $e^{it\q_1\q_2}$, each costing $n^2$ $C_z$ gates, totaling $O(n^2)$ elementary qubit gates.

The per-Fock discretization errors for the Gaussian operations above are shown in \autoref{fig:gate-errors}: rotation error $\epsilon_{\RR{\pi/4}}$ in (b), squeezing error $\epsilon_{\SS{1.0}}$ in (c), and beam splitter error $\epsilon_{\BS{\pi/2}}$ in (d). Each plot shows the error when the gate is applied to individual Fock states $\ketf{k}$, all exhibiting similar exponential scaling consistent with the QFT error analysis.


Similarly, $e^{it\q_j\q_k}$ has zero discretization error on position encoding (and $e^{it\p_j\p_k}$ has zero error on momentum encoding), following the same principle as \autoref{eq:zero-err}. The total unitary error is bounded by the sum of QFT errors at each basis-switching step.

\section{Hybrid Oscillator-Qubit Gates}\label{sec:hybrid}

In this section, we extend our framework to simulate conditional operations where a qubit controls a bosonic operation. The key principle remains the same: position/momentum-only operations are exact on their respective encodings, and errors arise only from QFT basis switching. The conditional versions of Gaussian gates follow the same decomposition structure and error analysis as their unconditional counterparts.

\subsection{Conditional Displacement}

On hybrid CV-DV systems, the conditional displacement gate on one qubit and one mode is defined as,
\begin{equation}
    \CD{\alpha} = e^{Z (\alpha\a^\dag-\alpha^*\a)},
\end{equation}
where $Z$ is the Pauli-Z operator on the control qubit.

To decompose conditional displacement, we first introduce the conditional parity gate $\CP$, defined as
\begin{equation}\label{eq:def-cp}
    \CP = \dyad{0} \otimes I + \dyad{1} \otimes \Par,
\end{equation}
where $\Par = e^{i\pi\n}$ is the parity operator that acts on the position wave function as $\Par: \psi(q) \mapsto \psi(-q)$.
The key property of the parity operator is that it conjugates displacement:
\begin{equation}\label{eq:parity-conj}
    \Par \D{\alpha} \Par = \D{-\alpha}.
\end{equation}
The discretization of the conditional parity gate is straightforward.
Since $\tPar = \prod_{j=0}^{n-1} X_j$ (Pauli-X on each qubit of the mode register), the conditional parity gate is discretized as
\begin{equation}\label{eq:dv-cp}
    \tCP = \prod_{j=0}^{n-1} \mathrm{CNOT}_{c,j},
\end{equation}
where $\mathrm{CNOT}_{c,j}$ is the CNOT gate with the control qubit $c$ and target on the $j$-th qubit of the mode register.
The circuit is shown in \autoref{fig:dv-cp}.
Note that the discretized conditional parity gate requires no QFT and thus introduces zero simulation error.


Using the conditional parity gate, the conditional displacement can be decomposed as
\begin{equation}\label{eq:cd-decomp}
\begin{aligned}
    \CD{\alpha} = &
    \dyad{0} \otimes \D{\alpha} + \dyad{1} \otimes \D{-\alpha}
    \\ = &
    \CP \cdot \D{\alpha} \cdot \CP,
\end{aligned}
\end{equation}
where the second equality follows from \autoref{eq:parity-conj} and the definition of $\CP$ in \autoref{eq:def-cp}.
This decomposition shows that the conditional displacement is simply an unconditional displacement sandwiched between two conditional parity gates.

Therefore, the discretization of the conditional displacement gate $\CD{\alpha}$ on position-encoding qubits is,
\begin{equation}\label{eq:dv-cd}
\begin{aligned}
    &
    \encq\pp{\CD{\alpha}}
    \\ = &
    \tCP \cdot \encq\pp{\D{\alpha}} \cdot \tCP
    \\ = &
    \tCP \cdot \encq\pp{e^{i\sqrt{2}(\Re\alpha) \q}} \cdot \tF^\dag 
    \\ & \cdot \encp\pp{e^{-i\sqrt{2}(\Im\alpha) \p}} \cdot \tF \cdot \tCP,
\end{aligned}
\end{equation}
where the displacement decomposition follows from \autoref{eq:dv-disp}.

Since the conditional parity gates require no QFT, the conditional displacement uses 2 QFTs, giving
\begin{equation}\label{eq:c-disp-err-bound}
    \epsilon_{\CD{\alpha}}(\ket{\psi}) \lesssim 2\eps_F.
\end{equation}
The gate cost of $\CD{\alpha}$ is 2 QFTs plus $2n$ $R_z$ gates and $2n$ CNOT gates (two $\tCP$ gates), totaling $O(n^2)$ elementary qubit gates.

\subsection{Conditional Gaussian Operations with Quadratic Hamiltonians}\label{sec:c-gaussian}

The conditional Gaussian operations are defined as controlled versions of the standard Gaussian gates.
The conditional rotation is
\begin{equation}\label{eq:c-rotation}
    \CR{\theta} = e^{-i\theta Z(\n+\hf{1})},
\end{equation}
the conditional squeezing is
\begin{equation}\label{eq:c-squeezing}
    \CS{r} = e^{\hf{r}Z(\a^2-\a^{\dag 2})},
\end{equation}
and the conditional beam splitter is
\begin{equation}\label{eq:c-beamsplitter}
    \CBS{\theta} = e^{-i\hf{\theta} Z(\ad\b+\a\bd)}.
\end{equation}

To simulate these, we need conditional versions of the elementary Gaussian gates from \autoref{thm:basis-cv}.
The conditional elementary Gaussian gate $e^{it Z\q_j\q_k}(j\ne k)$ can be simulated as,
\begin{equation}\label{eq:enc-cejk}
\begin{aligned}
    &
    \encq\pp{e^{it Z\q_j\q_k}}
    \\=&
    \prod_{r=0}^{n-1} \prod_{s=0}^{n-1} e^{i2^{r+s-1}\lambda^2 t Z Z_{j,r} Z_{k,s}}
    \\=&
    \prod_{r=0}^{n-1}
    \pp{
        \CNOT_{j,r}
        \prod_{s=0}^{n-1} e^{i2^{r+s-1}\lambda^2 t Z Z_{k,s}}
        \CNOT_{j,r}
    },
\end{aligned}
\end{equation}
where $\CNOT_{j,r}$ is the CNOT gate on the control qubit and the $r$-th qubit in the $j$-th register.
The gate count is $2n$ CNOT and $n^2$ $C_z$ gates when $j\ne k$.
When $j=k$,
\begin{equation}\label{eq:enc-cejj}
\begin{aligned}
    &
    \encq\pp{e^{it Z\q_j^2}}
    \\=&
    \prod_{0\le r < s \le n-1} e^{i2^{r+s}\lambda^2 t Z Z_{j,r} Z_{j,s}}
    \\=&
    \prod_{r=0}^{n-1}
    \pp{
        \CNOT_{j,r}
        \prod_{s=r+1}^{n-1} e^{i2^{r+s}\lambda^2 t Z Z_{j,s}}
        \CNOT_{j,r}
    },
\end{aligned}
\end{equation}
and the gate count is reduced to $2n$ CNOT and $\hf{n(n-1)}$ $C_z$ gates.
The conditional $e^{it\q_j\p_k}$ and $e^{it\p_j\p_k}$ can be encoded similarly on the corresponding basis.

Using these conditional elementary gates together with the decomposition principle for unconditional Gaussian gates, the conditional rotation on position-encoding qubits is:
\begin{equation}\label{eq:dv-cr}
\begin{aligned}
    &
    \encq\pp{\CR{\theta}}
    \\ =&
    e^{-\hf{i}Z\tan\hfth\Q^2} e^{-\hf{i}Z\sin\theta\P^2} e^{-\hf{i}Z\tan\hfth\Q^2}
    \\ =&
    \pp{
        \prod_{0\le s < t < n} e^{-i2^{s+t-1}\lambda^2 Z\tan\hfth Z_s Z_t}
    }
    \\ &
    \tF^\dag
    \pp{
        \prod_{0\le s < t < n} e^{i2^{s+t}\lambda^2 Z\sin\theta Z_s Z_t}
    }
    \tF
    \\ &
    \pp{
        \prod_{0\le s < t < n} e^{-i2^{s+t-1}\lambda^2 Z\tan\hfth Z_s Z_t}
    }
\end{aligned}
\end{equation}
when $\abs{\theta}\le\frac{\pi}{4}$, and
\begin{equation}\label{eq:dv-cr-general}
\begin{aligned}
    &
    \encq\pp{\CR{\theta}} = \encq\pp{\CR{\theta_0}} \encq\pp{\CR{\theta-\theta_0}},
\end{aligned}
\end{equation}
where $\theta_0\in\hf{\pi}\Z$ is chosen such that $\abs{\theta-\theta_0}\le\frac{\pi}{4}$, and
\begin{equation}\label{eq:dv-large-cr}
\begin{aligned}
    \encq\pp{\CR{\hf{\pi}}} =& \dyad{0} \otimes \tF + \dyad{1} \otimes \tF^\dag = \tF \cdot \tCP,
    \\
    \encq\pp{\CR{\pi}} =& \tCP,
    \\
    \encq\pp{\CR{-\hf{\pi}}} =& \dyad{0} \otimes \tF^\dag + \dyad{1} \otimes \tF = \tF^\dag \cdot \tCP.
\end{aligned}
\end{equation}

The conditional squeezing can be decomposed as,
\begin{equation}\label{eq:cs-decomp}
\begin{aligned}
    &
    \CS{r}
    \\ = &
    \dyad{0} \otimes \SS{r} + \dyad{1} \otimes \SS{-r}
    \\ = &
    \dyad{0} \otimes \pp{
        e^{i\hf{\mu_4(r)}\Q^2}
        e^{i\hf{\mu_3(r)}\P^2}
        e^{i\hf{\mu_2(r)}\Q^2}
        e^{i\hf{\mu_1(r)}\P^2}
    }
    \\ & +
    \dyad{1} \otimes \pp{
        e^{i\hf{\mu_4(-r)}\Q^2}
        e^{i\hf{\mu_3(-r)}\P^2}
        e^{i\hf{\mu_2(-r)}\Q^2}
        e^{i\hf{\mu_1(-r)}\P^2}
    }
    \\ = &
    e^{i\hf{(\mu_4(r)+\mu_4(-r))}\Q^2} e^{i\hf{(\mu_4(r)-\mu_4(-r))}Z\Q^2}
    \\ &
    e^{i\hf{(\mu_3(r)+\mu_3(-r))}\P^2} e^{i\hf{(\mu_3(r)-\mu_3(-r))}Z\P^2}
    \\ &
    e^{i\hf{(\mu_2(r)+\mu_2(-r))}\Q^2} e^{i\hf{(\mu_2(r)-\mu_2(-r))}Z\Q^2}
    \\ &
    e^{i\hf{(\mu_1(r)+\mu_1(-r))}\P^2} e^{i\hf{(\mu_1(r)-\mu_1(-r))}Z\P^2},
\end{aligned}
\end{equation}
where the coefficients are given in \autoref{eq:mu-coeffs}.

Applying the encoding to the result from \autoref{eq:cs-decomp}, the conditional squeezing on position-encoding qubits is:
\begin{equation}\label{eq:dv-cs}
\begin{aligned}
    &
    \encq\pp{\CS{r}}
    \\ =&
    e^{i\hf{(\mu_4(r)+\mu_4(-r))}\Q^2} e^{i\hf{(\mu_4(r)-\mu_4(-r))}Z\Q^2}
    \\ &
    \tF^\dag
    \pp{
        e^{i\hf{(\mu_3(r)+\mu_3(-r))}\Q^2} e^{i\hf{(\mu_3(r)-\mu_3(-r))}Z\Q^2}
    }
    \tF
    \\ &
    e^{i\hf{(\mu_2(r)+\mu_2(-r))}\Q^2} e^{i\hf{(\mu_2(r)-\mu_2(-r))}Z\Q^2}
    \\ &
    \tF^\dag
    \pp{
        e^{i\hf{(\mu_1(r)+\mu_1(-r))}\Q^2} e^{i\hf{(\mu_1(r)-\mu_1(-r))}Z\Q^2}
    }
    \tF,
\end{aligned}
\end{equation}
where the coefficients are given in \autoref{eq:mu-coeffs}.

The conditional beam splitter on position-encoding qubits is:
\begin{equation}\label{eq:dv-cbs}
\begin{aligned}
    &
    \encq\pp{\CBS{\theta}}
    \\ =&
    e^{-\hf{i}Z\tan\qtth(\Q_1\Q_2)}
    e^{-\hf{i}Z\sin\hfth(\P_1\P_2)}
    \\ &
    e^{-\hf{i}Z\tan\qtth(\Q_1\Q_2)}
    \\ =&
    \pp{
        \prod_{r,s=0}^{n-1} e^{-i2^{r+s-1}\lambda^2 Z\tan\qtth Z_{1,r} Z_{2,s}}
    }
    \\ &
    \tF_1^\dag \tF_2^\dag
    \pp{
        \prod_{r,s=0}^{n-1} e^{-i2^{r+s-1}\lambda^2 Z\sin\hfth Z_{1,r} Z_{2,s}}
    }
    \tF_1 \tF_2
    \\ &
    \pp{
        \prod_{r,s=0}^{n-1} e^{-i2^{r+s-1}\lambda^2 Z\tan\qtth Z_{1,r} Z_{2,s}}
    },
\end{aligned}
\end{equation}
for $\abs{\theta}\le\frac{\pi}{2}$.
For other angles, one can decompose the conditional beam splitter as $\CBS{\theta}=\CBS{\theta_0}\CBS{\theta-\theta_0}$, where $\theta_0\in\pi\Z$ and $\abs{\theta-\theta_0}\le\frac{\pi}{2}$.
The encoding of the large-angle conditional beam splitter $\CBS{\theta_0}$ is given by,
\begin{equation}\label{eq:dv-cbs-theta0}
\begin{aligned}
    &\encq\pp{\CBS{\pi}} = \tF_1 \tF_2 \cdot \tCP_1 \tCP_2 \cdot \tSWAP,
    \\
    &\encq\pp{\CBS{2\pi}} = \tPar_1 \tPar_2,
    \\
    &\encq\pp{\CBS{-\pi}} = \tF_1^\dag \tF_2^\dag  \cdot \tCP_1 \tCP_2\cdot \tSWAP,
\end{aligned}
\end{equation}
where $\tSWAP$ swaps the two mode registers, and $\tCP_j$ is the conditional parity gate on the control qubit and the $j$-th mode register.

The conditional Gaussian gates use the same number of QFTs as their unconditional counterparts, so the error bounds are unchanged.
\begin{equation}\label{eq:cr-err-bound}
    \eps_{\CR{\theta}}(\ket{\psi}) \lesssim 2\eps_F.
\end{equation}
The gate cost of $\CR{\theta}$ is 2 QFTs plus 3 conditional shearing layers $e^{itZ\q^2}$, each costing $\frac{n(n-1)}{2}$ $C_z$ and $2n$ CNOT gates, totaling $O(n^2)$ elementary qubit gates.
\begin{equation}\label{eq:cs-err-bound}
    \eps_{\CS{r}}(\ket{\psi}) \lesssim 4\eps_F.
\end{equation}
The gate cost of $\CS{r}$ is 4 QFTs plus 4 unconditional shearing layers $e^{it\q^2}$ and 4 conditional shearing layers $e^{itZ\q^2}$, totaling $O(n^2)$ elementary qubit gates.
\begin{equation}\label{eq:cbs-err-bound}
    \eps_{\CBS{\theta}}(\ket{\psi}) \lesssim 4\eps_F.
\end{equation}
The gate cost of $\CBS{\theta}$ is 4 QFTs plus 3 conditional shearing layers $e^{itZ\q_1\q_2}$, each costing $n^2$ $C_z$ and $2n$ CNOT gates, totaling $O(n^2)$ elementary qubit gates.

\section{Measurement Gates}\label{sec:measurements}

Measurements in our framework are either exact or have error determined by the number of QFTs required for basis preparation. We cover three important measurement types: homodyne detection, heterodyne detection, and photon number counting.

\subsection{Homodyne Detection}

Homodyne detection measures a quadrature observable and projects the state onto a quadrature eigenstate.
For example, the position basis homodyne detection $\M$ on the quadrature $\q$ returns a random variable $x$ with probability density function (PDF) $\abs{\psi(x)}^2$.
Position basis homodyne is simulated by direct Pauli-Z measurement on the position-encoded qubits, while momentum basis homodyne requires one QFT followed by Pauli-Z measurement.
In both cases, the measurement outcome is given by 
\begin{equation}
\begin{aligned}
    x = \lambda \bb{\sum_{j=0}^{n-1} 2^j b_j - \hf{N-1}},
\end{aligned}
\end{equation}
 where $b_j \in \{0,1\}$ are the measurement bits.
Position basis homodyne is \emph{exact} ($\eps = 0$) on position encoding, while momentum basis homodyne has error $\eps_F$ from the single QFT.
For the general case where one measures several commuting quadratures simultaneously, one can expand them into a symplectic basis $\kk{\Q_1,\cdots,\Q_m;\P_1,\cdots,\P_m}$, find a Gaussian operator to transform them into the position basis, and then perform the position basis homodyne detection (Pauli-Z measurement).
The error is determined by the number of QFTs in the Gaussian transformation.

\subsection{Heterodyne Detection}

Heterodyne detection measures both quadratures $\Re\a$ and $\Im\a$ simultaneously, projecting the state onto a coherent state.
The measurement outcome has Gaussian uncertainty due to the uncertainty principle.
For a single-mode state $\ket{\psi}$, heterodyne detection returns a random sample $\alpha\in\C$ with probability density $\Pr(\alpha) = \pi^{-1} \abs{\cip{\alpha}{\psi}}^2$.

On CV systems, heterodyne detection can be implemented by adding one ancilla mode $\b$ initialized in vacuum, applying a 50:50 beam splitter $\BS{\frac{\pi}{2}}$ to perform the symplectic basis change
\begin{equation}\label{eq:heterodyne-basis}
\begin{gathered}
    \q_1 \mapsto \q_1' = \sqhf{\q_1+\p_2}, \quad
    \p_1 \mapsto \p_1' = \sqhf{\p_1-\q_2}, \\
    \q_2 \mapsto \q_2' = \sqhf{\q_2+\p_1}, \quad
    \p_2 \mapsto \p_2' = \sqhf{\p_2-\q_1},
\end{gathered}
\end{equation}
and then performing homodyne detection on the transformed quadratures $\q_1'$ and $\p_2'$.
On DV systems, this requires simulating the beam splitter $\BS{\frac{\pi}{2}}$, which costs 4 QFTs and $3n^2$ $C_z$ gates, followed by Pauli-Z measurement on the $2n$ qubits encoding both modes.
The circuit is shown in \autoref{fig:dv-heterodyne}.

\subsection{Photon Number Counting}

\begin{figure}
\centering
\subfloat[Heterodyne detection circuit\label{fig:dv-heterodyne}]{%
    \includegraphics[width=\linewidth]{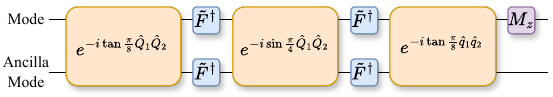}
}
\\
\subfloat[Photon number counting circuit\label{fig:dv-photon-counting}]{%
    \includegraphics[width=\linewidth]{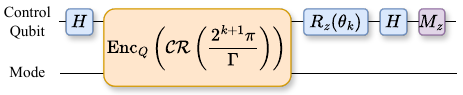}
}
\caption{
    Qubit circuits to simulate measurement protocols on position-encoding qubits.
    (a) Heterodyne detection measures both quadratures simultaneously by applying a beam splitter followed by homodyne detection on two qubit-simulated modes: the target mode and an ancilla mode initialized in vacuum.
    (b) Photon number counting uses iterative quantum phase estimation, running for $k=\gamma-1,\gamma-2,\cdots,0$ to extract all $\gamma$ bits of information. The feedback phase is $\theta_k = -\pi\sum_{\ell=k+1}^{\gamma-1} \frac{b_\ell}{2^{\ell-k}}$, computed from previous binary measurement results $\kk{b_\ell}_{\ell=k+1}^{\gamma-1}$.
}
\label{fig:dv-measurements}
\end{figure}

Photon number counting measures the number operator $\n = \ad\a$ and projects onto a Fock state.
For a CV mode $\sum_{k=0}^{\Gamma} \psi_k \ketf{k}$, it returns a random variable $k$ with probability $\abs{\psi_k}^2$, with Fock-level cutoff $\Gamma$.
Let $\gamma=\lceil\log_2(\Gamma+1)\rceil$ denote the number of bits needed to represent the photon number.
One can simulate photon number counting by quantum phase estimation with unitary $U=e^{i\frac{2\pi}{\Gamma}\n}$, which returns a phase angle $\frac{2\pi k}{\Gamma}$ with probability $\abs{\psi_k}^2$.
With \emph{iterative quantum phase estimation}, one needs only one ancilla qubit to extract all $\gamma$ bits of information iteratively.
The algorithm requires controlled-$U^{2^k}$ gates $e^{i\frac{2^{k+1}\pi}{\Gamma}Z\n}$ for $k=\gamma-1,\gamma-2,\cdots,0$, which are conditional rotations as discussed above.
The simulation cost is 1 conditional rotation gate, 2 Hadamard gates and 1 $R_z$ gate per bit of photon number.
The circuit is shown in \autoref{fig:dv-photon-counting}.
Each conditional rotation $\CR{\theta_k}$ uses 2 QFTs.

\section{Discussion}\label{sec:discussion}

We have presented a comprehensive framework for simulating hybrid CV-DV quantum processors on qubit-only systems using position encoding.
Our key results are as follows.
Any quadratic Gaussian operation can be decomposed into elementary gates $e^{it\q_j\q_k}$, $e^{it\p_j\p_k}$, $e^{it\q_j\p_k}$, each simulatable with $O(n^2)$ elementary qubit gates as shown in \autoref{thm:basis-cv} and \autoref{tab:basic-gates}.
Position operator-only unitaries are exact on position encoding, while momentum operator-only unitaries are exact on momentum encoding, with the only error source being QFT basis switching, as explained in \autoref{sec:enc}.
The per-Fock QFT error satisfies $\eps_F(\ketf{k}) \lesssim \exp\bb{(4.2125/\sqrt{N} + 0.1027)k - 5.7903\sqrt{N} + 16.2724}$ as characterized in \autoref{lem:qft_err_per_fock}, with the QFT count for each gate type summarized in \autoref{tab:gate-decomp}.
Assuming the QFT error of each intermediate state is bounded by $\eps_F$, the gate error is at most $k\eps_F$ where $k$ is the number of QFTs, as listed in \autoref{tab:gate-decomp}.
All gates in the phase-space instruction set (beam splitter, single-qubit rotation, conditional displacement), along with additional Gaussian and conditional Gaussian operations (displacement, squeezing, conditional squeezing, conditional rotation, conditional beam splitter), are efficiently simulable, as are all measurement protocols (homodyne, heterodyne, photon counting).

For practical resource estimation, given a target Fock level $\Gamma$ and precision $\epsilon$, \autoref{lem:qft_err_per_fock} requires finding $N = 2^n$ such that $\eps_F(\ketf{\Gamma}) \le \epsilon$:
\begin{equation}\label{eq:n-from-error}
    n \ge 2\log_2\pp{\frac{C + \sqrt{C^2 + 97.60\,\Gamma}}{11.5806}},
\end{equation}
where $C = 0.1027\,\Gamma + 16.2724 + \ln(1/\epsilon)$.
This follows from solving the quadratic inequality $5.7903 \times 2^n - C \times 2^{n/2} - 4.2125\,\Gamma \ge 0$.
Roughly, this gives $n=O(\log(\Gamma + \log(\epsilon\inv)))$ qubits per mode.
The gate count complexity is $O(n^2) = O(\log^2(\Gamma + \log(\epsilon\inv)))$ per hybrid gate.
This polylogarithmic per-gate scaling represents an \emph{exponential improvement} over Fock basis encoding methods, which face a fundamental complexity barrier: they require either $O(2^n n^2)$ quantum gates per gate through Trotter decomposition (as shown in \autoref{eq:fock-trotter-complexity} below), or exponential classical compilation time (as in the approach of \cite{arzani2025can}, \autoref{eq:arzani-compile}).
In contrast, our position encoding achieves $O(n^2)$ gate complexity per hybrid gate with explicit, efficiently computable decompositions requiring no expensive classical preprocessing.
The gate error is bounded by $k\eps_F$, independent of any Fock-level cutoff.
For states with bounded mean energy $E = \ev{\n}$, the dominant Fock contribution scales with $E$, and using $\Gamma = O(E/\epsilon^2)$ as a reference scale gives $n = O(\log(E/\epsilon^2 + \log(\epsilon\inv)))$ qubits per mode and gate count complexity $O(\log^2(E/\epsilon^2 + \log(\epsilon\inv)))$.

In this paper, we always encode CV modes into DV systems using position/momentum encoding, which makes Gaussian and conditional Gaussian gates relatively easy to simulate.
There are also other ways to encode CV modes into DV systems, such as the simple Fock basis encoding, $\sum_{k=0}^{\infty} c_k \ketf{k} \mapsto \frac{1}{\mathcal{N}} \sum_{k=0}^{\Gamma} c_k \ket{k}$, where $\Gamma$ is the Fock-level cutoff and $\mathcal{N} = \norm{\sum_{k=0}^{\Gamma} c_k \ket{k}}$ is the normalization factor.
In this case, the photon-number-related operations are easy to simulate, but the Gaussian operations become very expensive.
Refs.~\cite{liu2024hybrid} and \cite{crane2024hybrid} provide simulation of displacement ($\alpha\in\R$), beam splitter, and conditional beam splitter gates ($\phi=0$) on Fock basis encoding qubits.
In their approach, they split a displacement operator $e^{i\alpha(\a+\ad)}$ ($\alpha\in\R$) into two parts,
\begin{equation}
\begin{aligned}
    &
    \a + \ad
    \\ = &
    \bb{
        \pmat{0} + \oplus_{j=1}^{\infty} \pmat{
            0 & \sqrt{2j} \\
            \sqrt{2j} & 0
        }
    }
    \\ & +
    \bb{
        \oplus_{j=1}^{\infty} \pmat{
            0 & \sqrt{2j+1} \\
            \sqrt{2j+1} & 0
        }
    }
    \\ = &
    H_{\mathrm{even}} + H_{\mathrm{odd}},
\end{aligned}
\end{equation}
and use Trotter decomposition to turn it into a problem of simulating $e^{i\tau H_{\mathrm{even}}}$ and $e^{i\tau H_{\mathrm{odd}}}$, which requires Newton iteration to approximate the square root functions.
The number of CNOT gates in a single Trotter step using $n_{NI}$ Newton iterations is
\begin{equation}
    N_{CNOT} = (270n_{NI} + 126)n^2 + (144+228 n_{NI})n - 12 n_{NI} - 22,
\end{equation}
and the number of Trotter steps is,
\begin{equation}\label{eq:fock-trotter-complexity}
    N_{\exp} \leq \left\lceil
        3 N \alpha \pp{\frac{25}{3}}^{k} \pp{\frac{2\alpha N}{\eps}}^{1/(2k)}
    \right\rceil,
\end{equation}
in the worst case with the Trotter-Suzuki formula of order $2k$ to ensure $\hf{\eps}$-small error.
Even with constant $n_{NI}$, the total complexity scales as $O(2^n n^2)$---exponential in the number of qubits---and involves multiple error sources including Trotter and Newton iterations that are difficult to accurately estimate.
Similarly, they can simulate a beam splitter or conditional beam splitter by splitting the Hamiltonian into 4 parts.
In comparison, our method can simulate displacement, beam splitter, and conditional beam splitter with $O(n^2)$ qubit gates and error given by \autoref{eq:qft-err-numerical}, due to the simple matrix form of those unitaries in position encoding.

\begin{figure}[t]
    \centering
    \includegraphics[width=.95\linewidth]{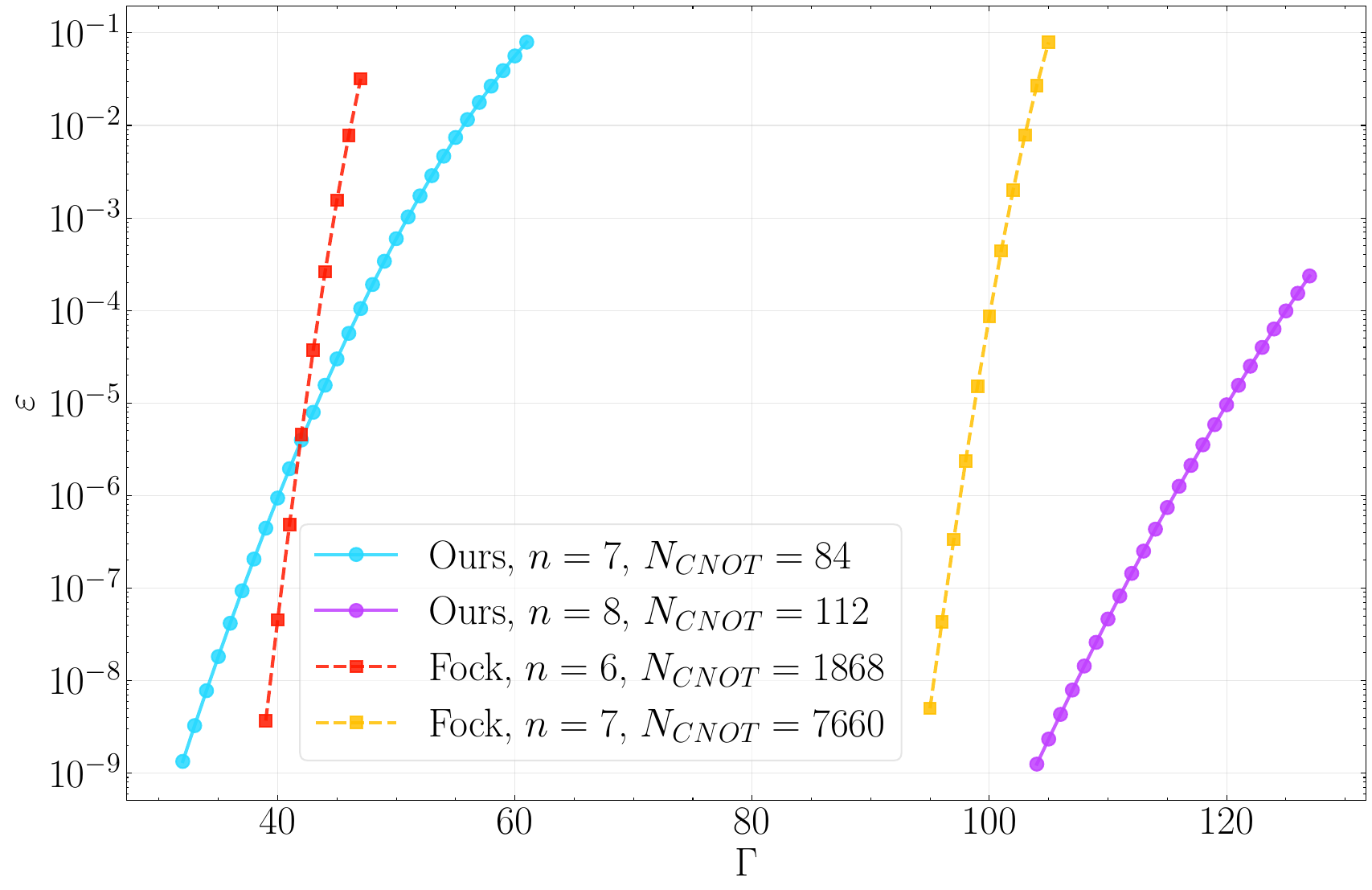}
    \caption{
        Comparison of displacement operator simulation error $\epsilon$ versus Fock-level cutoff $\Gamma$ for wave function (WF) encoding and Fock basis encoding with $\alpha=2.0$.
        WF encoding (solid lines) achieves comparable accuracy to Fock encoding (dashed lines) with significantly fewer CNOT gates.
        At $n=7$ qubits, WF encoding uses 84 CNOTs while Fock encoding requires 7660 CNOTs after Qiskit transpilation---a $91\times$ reduction at equal qubit count.
    }
    \label{fig:compare-fock}
\end{figure}

To illustrate this concretely, \autoref{fig:compare-fock} compares the displacement operator simulation error for position encoding and Fock basis encoding as a function of Fock-level cutoff $\Gamma$, for displacement parameter $\alpha=2.0$.
The Fock basis circuit is constructed from the exact $2^n \times 2^n$ displacement unitary and transpiled using Qiskit with optimization level 1 and basis gates $\{U_3, \mathrm{CNOT}\}$.
For instance, position encoding with $n=7$ qubits requires only 84 CNOT gates yet achieves a comparable error level to Fock basis encoding with $n=6$ qubits, which needs 1868 CNOT gates---a $22\times$ reduction in gate count for equivalent accuracy. At the same qubit count $n=7$, Fock basis encoding requires 7660 CNOT gates versus 84 for position encoding---a $91\times$ reduction. At $n=8$, position encoding achieves lower error than Fock basis encoding at $n=7$ while still requiring only 112 CNOT gates versus 7660---a $68\times$ reduction despite improved accuracy.
This numerical comparison confirms the exponential separation in gate complexity predicted by the theoretical analysis.

Another related work~\cite{arzani2025can} uses the Solovay-Kitaev theorem to show that any oscillator unitary $U$ can be simulated with a unitary $V$ on an $n$-qubit system where
\begin{equation}\label{eq:arzani-dim}
\begin{aligned}
    N = 2^n = O\left(
        \frac{E}{\eps^4} E_U\pp{
            \frac{64 E}{\eps^2}
        }
    \right),
\end{aligned}
\end{equation}
where $E_U$ is a non-decreasing $U$-dependent function, $E$ is the mean energy constraint and $\eps$ is the trace distance error, which is of the same order as our Euclidean norm error.
Moreover, the classical resource needed to compile the qubit circuit is,
\begin{equation}\label{eq:arzani-compile}
    O\left(
        \frac{NE^2}{\eps^4}
    \right)
    =
    O\left(
        \frac{E^3}{\eps^8}
        E_{U} \pp{
            \frac{64E}{\eps^2}
        }
    \right),
\end{equation}
which scales exponentially with the number of qubits $n$ since $N = 2^n$.
While this approach applies to arbitrary oscillator unitaries, the classical compilation overhead renders it impractical for large-scale simulations.
In contrast, our method focuses on gates with at most quadratic Hamiltonians, which enables explicit gate decompositions with $O(n^2)$ complexity and no expensive classical preprocessing.

For simulating arbitrary bosonic Hamiltonians, if one can decompose them into steps comprising only gates with at most quadratic Hamiltonians using Trotter, the BCH formula, or similar techniques with polynomial overhead~\cite{crane2024hybrid,kang2023leveraging}, then our method retains its exponential advantage.
The fundamental complexity separation is clear: existing Fock basis approaches require exponential resources in $n$---either quantum or classical---whereas our position encoding achieves $O(n^2)$ gate complexity with efficient classical compilation.

\begin{figure}[t]
    \centering
    \includegraphics[width=.95\linewidth]{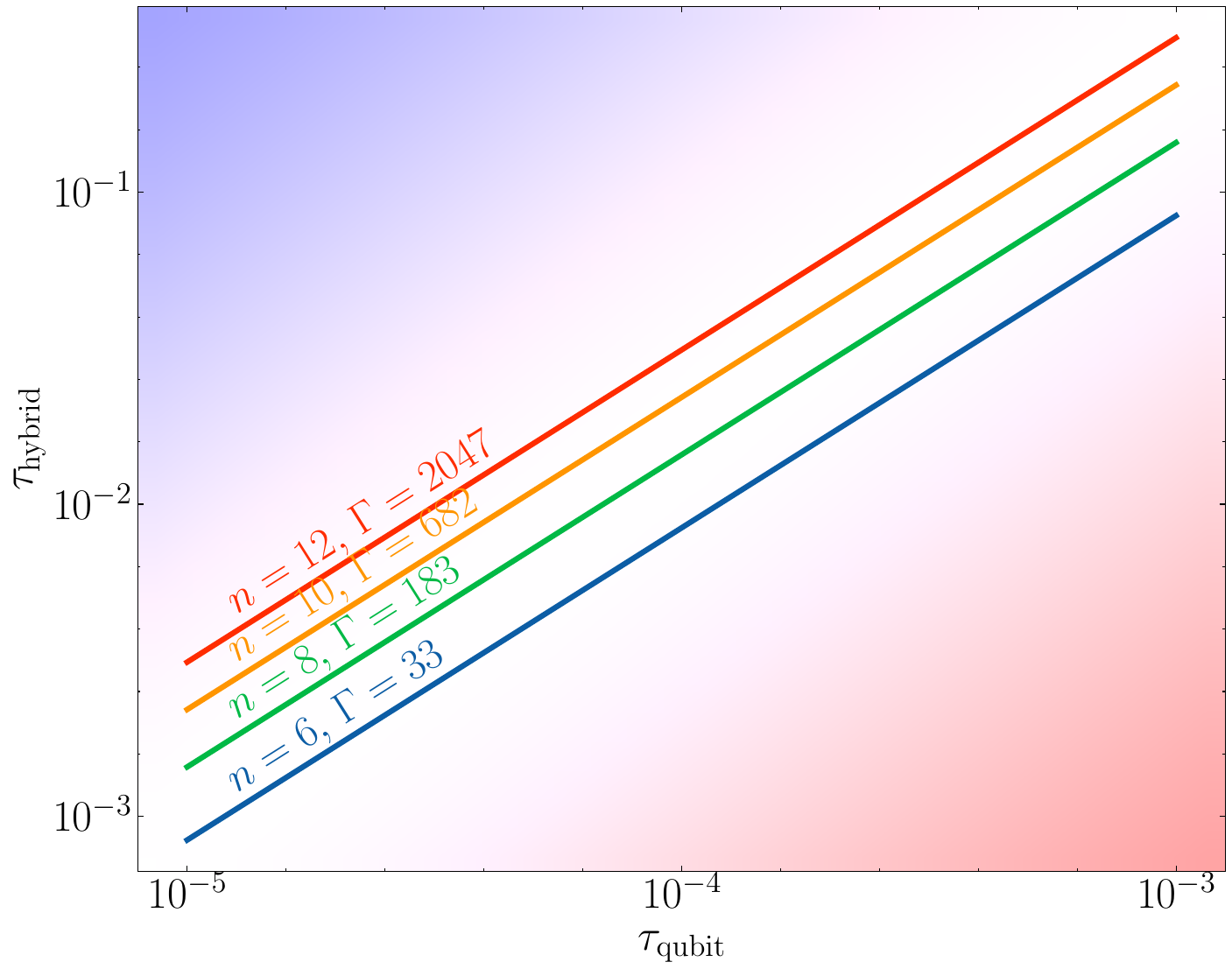}
    \caption{
        Potential qubit or hybrid CV-DV advantage for fixed simulation error $\epsilon=10^{-4}$.
        For each number of qubits per mode $n$, the Fock-level cutoff $\Gamma$ is chosen to achieve the target error using \autoref{lem:qft_err_per_fock}.
        The axes represent normalized gate operation times
            $\tau_{\mathrm{hybrid}}=t_{\mathrm{hybrid}}/T_{2,\mathrm{hybrid}}$,
            $\tau_{\mathrm{qubit}}=t_{\mathrm{qubit}}/T_{2,\mathrm{qubit}}$,
        where $t$ is the gate operation time and $T_2$ is the decoherence time.
        The blue region indicates qubit advantage, while the red region indicates hybrid CV-DV advantage, separated by the lines $\tau_{\mathrm{hybrid}} = (2n^2+2n)\tau_{\mathrm{qubit}}$,
        where $2n^2+2n$ is the CNOT gate count for simulating one conditional displacement.
    }
    \label{fig:advantage}
\end{figure}

Finally, we discuss the advantage of simulating hybrid CV-DV processors with qubits.
With near-term hardware, the gate count is limited when taking decoherence time and gate operation time into consideration.
As a benchmark, we consider a hybrid CV-DV circuit consisting of conditional displacement gates only, and compare it with a qubit circuit that simulates it using the method in this paper with $n$ qubits per CV mode.
Consider the most expensive case where each displacement parameter is a general complex number, then one needs 2 QFT and $2n$ $C_z$ gates for one conditional displacement on qubits, which costs $2n^2+2n$ CNOT gates in total.
Let $t_{\mathrm{hybrid}}$ and $t_{\mathrm{qubit}}$ be the gate operation time for one conditional displacement gate on hybrid CV-DV systems and one CNOT gate on qubit systems, and $T_{2,\mathrm{hybrid}}$ and $T_{2,\mathrm{qubit}}$ be the decoherence time for hybrid CV-DV systems and qubit systems, respectively.
The normalized gate operation times $\tau_{\mathrm{hybrid}}=t_{\mathrm{hybrid}}/T_{2,\mathrm{hybrid}}$ and $\tau_{\mathrm{qubit}}=t_{\mathrm{qubit}}/T_{2,\mathrm{qubit}}$ are indicators of their computational power, as the system with smaller $\tau$ can perform more gates within the decoherence time.
We compare the hybrid or qubit advantage in \autoref{fig:advantage} for fixed simulation error $\epsilon=10^{-4}$.
For each value of $n$, the Fock-level cutoff $\Gamma$ is determined from \autoref{lem:qft_err_per_fock} to achieve the target error, resulting in trade-offs between quantum resource ($n$) and classical simulation complexity ($\Gamma$).

\section{Conclusion}\label{sec:conclusion}

We have introduced a position encoding framework for simulating hybrid oscillator-qubit quantum processors on qubit-only systems, achieving $O(\log^2(\Gamma + \log(\epsilon\inv)))$ elementary qubit gates per hybrid gate for all operations with at most quadratic Hamiltonians.
This encompasses the phase-space instruction set (beam splitter, single-qubit rotation, conditional displacement) and extends to all Gaussian and conditional Gaussian operations including squeezing, conditional squeezing, conditional rotation, and conditional beam splitter.
The polylogarithmic per-gate complexity represents an exponential improvement over Fock basis encoding approaches, which require either exponential quantum gates through Trotter decomposition or exponential classical compilation time.

Beyond establishing polynomial overhead for hybrid algorithms on qubit processors, this work provides several practical contributions: (1) explicit gate decompositions with rigorous error bounds for resource estimation, (2) systematic numerical characterization of QFT errors enabling precise implementation planning, and (3) efficient measurement protocols including homodyne, heterodyne, and photon number detection.
These results advance our understanding of computational power relationships between quantum architectures and provide concrete pathways for near-term implementations.

Several promising directions emerge from this framework.
First, comparative resource analysis for specific applications---such as quantum error correction with bosonic codes, quantum simulation of bosonic/fermionic systems, and quantum signal processing---would reveal when hybrid architectures provide practical advantages.
Second, extending these techniques to fault-tolerant settings could enable error-corrected implementations on both architectures.
Third, investigating optimal state preparation and measurement strategies within this encoding could further reduce resource requirements.
Finally, exploring connections to quantum signal processing and quantum machine learning applications may uncover new algorithmic opportunities enabled by efficient hybrid-to-qubit compilation.

\begin{acknowledgments}
    This work was supported by the U.S. Department of Energy, Office of Science, Advanced Scientific Computing Research, under contract number DE-SC0025384. 
\end{acknowledgments}

\section*{Data and Code Availability}

The numerical experiments and error analysis presented in this paper are reproducible using the code available at \url{https://github.com/helloluxi/cuda-cvdv}, which also includes a fully functional CUDA-based classical simulator for hybrid CV-DV quantum processors.

\appendix

\section{Gate Decompositions}\label{app:decomp}

\subsection{Rotation Gate}\label{app:decomp-rot}

Using the elementary decomposition,
\begin{equation}
    \begin{pmatrix}
        \cos\theta & \sin\theta \\
        -\sin\theta & \cos\theta
    \end{pmatrix}
    =
    \begin{pmatrix}
        1 & 0 \\
        -\tan\hfth & 1
    \end{pmatrix}
    \begin{pmatrix}
        1 & \sin\theta \\
        0 & 1
    \end{pmatrix}
    \begin{pmatrix}
        1 & 0 \\
        -\tan\hfth & 1
    \end{pmatrix},
\end{equation}
the rotation gate $\RR{\theta} := e^{-i\theta(\n+\hf{1})}$ can be decomposed as,
\begin{equation}\label{eq:rotation-decomp}
    \RR{\theta} = e^{-\hf{i}\tan\hfth\q^2} e^{-\hf{i}\sin\theta\p^2} e^{-\hf{i}\tan\hfth\q^2},
\end{equation}
an identity equivalent to \cite[Claim~2.1]{iyer2026efficient} applied to the $\mathfrak{sp}(2,\mathbb{R})$ algebra of the harmonic oscillator.
This requires 2 QFTs, $\frac{3n(n-1)}{2}$ $C_z$ gates.
We note that when $\theta>\hf{\pi}$, $\tan\hfth$ grows rapidly and could increase the QFT error of intermediate states.
One can split a large angle $\theta$ into an angle $\theta_0\in\hf{\pi}\Z$ and a small angle $\theta-\theta_0$, such that $\abs{\theta-\theta_0}\le\frac{\pi}{4}$, while rotation by $\theta_0$ itself is an elementary operator,
\begin{equation}\label{eq:elem-rotation}
\begin{aligned}
    \RR{\hf{\pi}} =& \F,
    \\
    \RR{\pi} =& \Par,
    \\
    \RR{-\hf{\pi}} =& \F\inv.
\end{aligned}
\end{equation}

\subsection{Squeezing Gate}\label{app:decomp-sq}

Using the elementary decomposition,
\begin{equation}
    \begin{pmatrix}
        e^{-r} & 0 \\
        0 & e^{r}
    \end{pmatrix}
    =
    \begin{pmatrix}
        1 & 0 \\
        -\frac{e^r - 1}{te^{-r}} & 1
    \end{pmatrix}
    \begin{pmatrix}
        1 & -te^{-r} \\
        0 & 1
    \end{pmatrix}
    \begin{pmatrix}
        1 & 0 \\
        \frac{e^r - 1}{t} & 1
    \end{pmatrix}
    \begin{pmatrix}
        1 & t \\
        0 & 1
    \end{pmatrix},
\end{equation}
the squeezing gate $\SS{r} := e^{\hf{r}(\a^2-\a^{\dag2})}$ can be decomposed as,
\begin{equation}\label{eq:squeeze-decomp}
    \SS{r} =
    e^{i\frac{1-e^{r}}{2te^{-r}}\q^2}
    e^{i\frac{te^{-r}}{2}\p^2}
    e^{i\frac{e^{r}-1}{2t}\q^2}
    e^{-i\frac{t}{2}\p^2},
\end{equation}
which requires 4 QFTs and $2n(n-1)$ $C_z$ gates.
We choose $t=e^{r/2}\sqrt{\abs{e^{r}-1}}$ to balance the two groupings in the sum of shearing magnitudes:
\begin{equation}\label{eq:squeezing-min-coeff}
\begin{aligned}
    &
    \abs{t} + \abs{\frac{e^{r}-1}{t}} + \abs{te^{-r}} + \abs{\frac{e^{r}-1}{te^{-r}}}
    \\=&
    \abs{t} (1+e^{-r}) + \abs{\frac{e^{r}-1}{t}} (1+e^{r}).
\end{aligned}
\end{equation}
This gives the shearing coefficients in \autoref{eq:mu-coeffs}.

\subsection{Beam Splitter Gate}\label{app:decomp-bs}

Finally, since
\begin{widetext}
\begin{equation}
    \begin{pmatrix}
        \cos\hfth & 0 & 0 & \sin\hfth \\
        0 & \cos\hfth & \sin\hfth & 0 \\
        0 & -\sin\hfth & \cos\hfth & 0 \\
        -\sin\hfth & 0 & 0 & \cos\hfth
    \end{pmatrix}
    =
    \begin{pmatrix}
        1 & 0 & 0 & 0 \\
        0 & 1 & 0 & 0 \\
        0 & -\tan\qtth & 1 & 0 \\
        -\tan\qtth & 0 & 0 & 1
    \end{pmatrix}
    \begin{pmatrix}
        1 & 0 & 0 & \sin\hfth \\
        0 & 1 & \sin\hfth & 0 \\
        0 & 0 & 1 & 0 \\
        0 & 0 & 0 & 1
    \end{pmatrix}
    \begin{pmatrix}
        1 & 0 & 0 & 0 \\
        0 & 1 & 0 & 0 \\
        0 & -\tan\qtth & 1 & 0 \\
        -\tan\qtth & 0 & 0 & 1
    \end{pmatrix},
\end{equation}
\end{widetext}
the beam splitter $\BS{\theta} := e^{-i\hfth(\ad\b+\a\bd)}$ can be decomposed as,
\begin{equation}\label{eq:bs-decomp-simple}
    \BS{\theta} =
    e^{-i\tan\qtth\q_1\q_2}
    e^{-i\sin\hfth\p_1\p_2}
    e^{-i\tan\qtth\q_1\q_2},
\end{equation}
which uses 4 QFTs and $3n^2$ $C_z$ gates.
Similar to the rotation gate, for a large rotation angle $\theta$, one can split it into an angle $\theta_0\in\pi\Z$ plus a small angle $\theta-\theta_0$ such that $\abs{\theta-\theta_0}\le\hf\pi$, and
\begin{equation}\label{eq:bs-shortcut}
\begin{aligned}
    \BS{\pi} =& \F_1 \F_2 \SWAP,
    \\
    \BS{2\pi} =& \Par_1 \Par_2,
    \\
    \BS{-\pi} =& \F_1^\dag \F_2^\dag \SWAP.
\end{aligned}
\end{equation}


\bibliographystyle{unsrt}
\bibliography{ref.bib}

\end{document}